\documentclass{article}
\usepackage[utf8]{inputenc}
\usepackage[english]{babel}
\usepackage{csquotes}
\usepackage[backend=bibtex,citestyle=numeric-comp,sorting=none]{biblatex}
\usepackage{lineno,hyperref}
\usepackage{amssymb}
\usepackage{mathtools}

\usepackage[binary-units=true]{siunitx}
\usepackage{subfig}

\usepackage{cleveref} 
\usepackage{xcolor}
\usepackage{booktabs}
\usepackage{tabularx}
\usepackage{algorithm}
\usepackage[noend]{algpseudocode}
\def\BState{\State\hskip-\ALG@thistlm}
\algdef{SE}[DOWHILE]{Do}{doWhile}{\algorithmicdo}[1]{\algorithmicwhile\ #1}

\newcommand{\bokeh}{Bokeh}
\newcommand{\cxx}{C/C++}
\newcommand{\fortran}{Fortran}
\newcommand{\python}{Python}
\newcommand{\matlab}{MATLAB}
\newcommand{\paraprobe}{PARAPROBE}
\newcommand{\cgal}{CGAL}
\newcommand{\qhull}{QHull}
\newcommand{\voroxx}{Voro$++$}
\newcommand{\hdf}{HDF5}
\newcommand{\xdmf}{XDMF}
\newcommand{\hdmf}{HDF5/XDMF}
\newcommand{\omp}{OpenMP}

\newcommand{\imk}{Intel Math Kernel}

\newcommand{\ivas}{IVAS}

\newcommand{\rapidxml}{RapidXML}
\newcommand{\ashapes}{$\alpha$-shapes}
\newcommand{\ashape}{$\alpha$-shape}
\newcommand{\gshapes}{$\gamma$-shapes}

\newcommand{\althreesc}{Al${}_3$Sc}

\newcommand{\alscsi }[2]{Al-\SI[mode=text]{#1}{}Sc-\SI[mode=text]{#2}{}Si}
\newcommand{\itswtpercent}{(wt. \SI[mode=text]{}{\percent})}
\newcommand{\cameca}{CAMECA/AMETEK}

\newcommand{\maxsep}[4]{$d_{max} = [#1,#2,#3] \SI[mode=math]{}{\nano\meter}, N_{min} = \SI[mode=math]{#4}{}$}
\newcommand{\spatstat}[5]{$k = #1, r = [#2,#3,#4] \SI[mode=math]{}{\nano\meter}, d_{srf} \geq \SI[mode=math]{#5}{\nano\meter}$}
\newcommand{\spatstatnok}[4]{$r = [#1,#2,#3] \SI[mode=math]{}{\nano\meter}, d_{srf} \geq \SI[mode=math]{#4}{\nano\meter}$}  
\newcommand{\spatstatnor}[4]{$k = #1, r = [#2,#3,#4] \SI[mode=math]{}{\nano\meter}$}  

\newcommand{\ionsurf}[1]{$d_{prb} = \SI[mode=math]{#1}{\nano\meter}$}
\newcommand{\vorotess}[1]{$d_{ero} = \SI[mode=math]{#1}{\nano\meter}$}

\newcommand{\ashapeinit}[1]{$d_{bin} = \SI[mode=math]{#1}{\nano\meter}$}

\usepackage{authblk}

\title{On Strong Scaling and Open Source Tools for \\ Analyzing Atom Probe Tomography Data}
\author[1]{Markus K\"uhbach}
\author[1]{Priyanshu Bajaj}
\author[2]{Murat Han {\c{C}}elik} 
\author[3,1]{\\ Eric A. J\"agle}
\author[1,4]{Baptiste Gault}
\affil[1]{Max-Planck-Institut f\"ur Eisenforschung GmbH (MPIE), Max-Planck-Stra{\ss}e 1, D-40237 D\"usseldorf, Germany}
\affil[2]{Institute for Advanced Simulation (IAS), J\"ulich Supercomputing Centre (JSC), Wilhelm-Johnen-Stra{\ss}e, D-52425 J\"ulich, Germany}
\affil[3]{Universit\"at der Bundeswehr M\"unchen, Werner-Heisenberg-Weg 39, D-85577 Neubiberg, Germany}
\affil[4]{Department of Materials, Imperial College London, Royal School of Mines, London SW7 2AZ, United Kingdom}
\date{}

\begin{document}
\maketitle

\section{Abstract}
Atom probe tomography (APT) has matured to a versatile nanoanalytical characterization tool with applications that range from materials science to geology and possibly beyond. Already, well over 100 APT microscopes exist worldwide. Information from the APT data requires a post-processing of the reconstructed point cloud which is realized via basic implementations of data science methods, mostly executed with proprietary software. Limitations of the software have motivated the APT community to develop supplementary post-processing tools to cope with increasing method complexity and higher quality demands: examples are how to improve method transparency, how to support batch processing capabilities, and how to document more completely the methods and computational workflows to better align with the FAIR data stewardship principles.

One gap in the APT software tool landscape has been a collection of open tools which support scientific computing hardware. Here, we introduce \paraprobe, an open source, efficient tool for the scientific computing of APT data. We show how to process several computational geometry, spatial statistics, and clustering tasks performantly for datasets as large as two billion ions. Our parallelization efforts yield orders of magnitude performance gains and deliver batch processing capabilities. We contribute these tools in an effort to open up APT data mining and simplify it to make tools for rigorous quantification, sensitivity analyses, and cross-method benchmarking available to practitioners\footnote{Corresponding author: m.kuehbach@mpie.de}.

\section{Keywords}
Atom probe tomography, spatial statistics, maximum separation method, clustering, Voronoi tessellation, computational geometry, parallelization, data mining, additive manufacturing, \paraprobe{}

\section{Introduction}
Atom probe tomography (APT) is a destructive microscopy and microanalysis technique which allows the characterization of specific microstructural features with near-atomic resolution in three dimensions. Using either controlled laser or high voltage pulses superimposed on a DC high voltage, APT relies on the process of field evaporation to remove individual atoms from a needle-shaped specimen in the form of ions. These are collected by a position-sensitive time-resolved detector system \cite{Gault2012b,Larson2013b,Lefebvre2016,Miller1996}. The time-of-flight of each ion allows for elemental identification with isotopic resolution. The association of a range of mass-to-charge-state-ratio values to a single element is usually referred to as ranging \cite{Hudson2011,Haley2017}. Following elemental identification, a combination of a reverse projection and a sequential depth increment computation allows to reconstruct a point cloud; and thereby reveal the original atomic arrangement of the specimen \cite{Gault2011b}.

Improvements in instrumentation and experimental protocols in the past decade have made  multi-million, as well as for some materials even billion, ion datasets accessible. Inspecting for instance the joint database for all APT microscopes of the MPIE yields a list of 743 datasets that were measured between January, 2016 and February, 2020 which have all at least \SI{100e6}{} ions collected. Combined with other microscopy techniques, in particular transmission electron microscopy \cite{Herbig2018}, makes APT a uniquely powerful tool for advanced materials characterization. The range of applications spans fields as diverse as physical metallurgy \cite{Hono1999,Kuzmina2015}, geology and planetary chronology \cite{Valley2014,Piazolo2016,White2017,Saxey2018}, solar energy harvesting \cite{Cojocaru-Miredin2018}, biology \cite{Perea2016,Rusitzka2018}, or semiconductors \cite{Voyles2002,Barnes2018,Giddings2018}. Specimens in these fields probe from single-crystalline, single-phase chemistry to complex multinary polycrystals with ten or more elements or amorphous phases \cite{Kontis2018,Li2018,Gin2017,Sepehriamin2017}. The range of materials amenable to APT analysis will keep expanding in the coming years with new cryo-preparation and transfer protocols being explored \cite{Schreiber2018,Chang2019}.

Post-processing of these APT data is a critical step in every study. Examples include tasks like reconstructing a specimen from the time-of-flight detector hit sequence \cite{Gault2012b} or characterizing spatial statistics of the nanoscale composition \cite{Zhao2018}. It is common to compute concentration fields to superimpose iso-surfaces on these fields for visualization, and to evaluate concentration profiles \cite{Hellman2000b}. Other frequent tasks are to characterize second-phase precipitates using clustering algorithms \cite{Hyde2000,Stephenson2007,Zelenty2017,Ghamarian2019} or to reconstruct microstructural features in the specimen from the ion positions using computational geometry methods \cite{Haley2009,Felfer2013,Felfer2016}.


The \textit{de facto} near-monopolistic APT instrument landscape means that the Integrated Visualization and Analysis Software (\ivas{}), is an almost mandatory starting point for most practitioners \cite{Ulfig2017,Reinhard2019}. \ivas{} is necessary because the raw data of an APT measurement are stored in proprietary container files. In return, though, the software offers a functionally-rich front end solution with which practitioners can execute almost all of above characterization tasks. \ivas{} has limitations, though, with respect to the transparency of the algorithms it uses, the fact that these cannot easily be modified, and lacking support for scientific computing hardware which would allow for applying sophisticated batch processing strategies. This motivated efforts by the APT community to develop complementary scripts and software tools \cite{Stephenson2007,Boll2007,Moody2008,Moody2009b,Haley2009,Yao2010,Saxey2011,Ceguerra2013,Felfer2013,Felfer2016,APTTools2017,Massive2017,Keutgen2017,Haley2018b}. Most of these constitute proof-of-concept implementations of novel algorithms or patches of missing \ivas{} functionalities. With a prime focus on serving as supplementary tools, though, these usually only provide for sequential execution.

This status quo of APT software tools is embedded in a global picture, with recent trends and new technological developments on the horizon. It is our opinion that these demand additional strategic considerations and pose new challenges for analyzing APT experiments with future-proof computational materials science data analysis methods:

\begin{itemize}
\item Current experiments lead to the collection of larger datasets, thanks to a wider field of view, a higher detection efficiency as well as the increase in yield provided by laser pulsing capabilities. 
\item Stronger quality demands to the analyses and increasingly more complex getting methods are a reality as well in APT.
\item Many individuals in the APT community see value in opening up software and file formats in an effort to improve on the documentation of the existing software. In addition, they also see value in reporting more detail about the data acquisition, the post-processing methods, and the workflows used in an effort to economize and improve the research process. This is in parts overseen by the Technical Committee of the International Field Emission Society.
\item With the stronger permeation of machine learning and artificial intelligence methods, one may argue that missing documentation or undisclosed data reduce the speed at which new data analysis techniques can be developed, tested for their effectiveness, and broadly deployed. 
\item Journals and funding agencies are likely to start enforcing stricter quality demands with respect to the curation of experimental data, including APT data.
\end{itemize}

One solution to cope with above challenges is to improve the documentation and curation of APT data as comprehensively and as automatically as possible. This aligns with the goals of the FAIR research and data stewardship principles \cite{Wilkinson2016,Draxl2020}. The acronym FAIR stands for research which is findable, accessible, interoperable, and reproducible. Only concerted efforts across the community could bring APT research closer to become compliant with the FAIR principles. This  will be rewarding because methods from scientific computing can be better utilized and with this especially those manual procedures reduced which are prone to user errors. Examples for this are the application of artificial intelligence tools, high-throughput analyses, scientific visualization, and wizards for automated report writing. 

\begin{figure}[!ht]
\centering
	\includegraphics[width=1.00\textwidth]{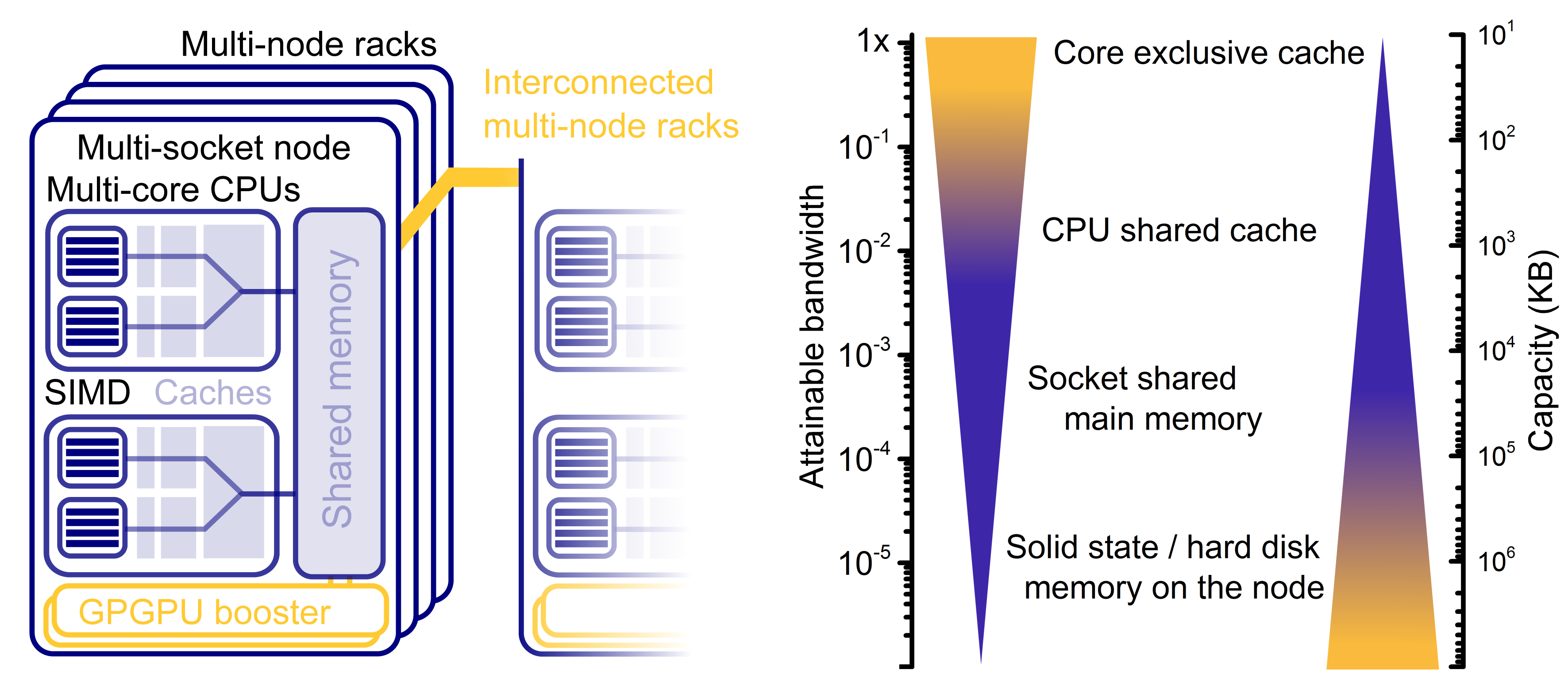}
\caption{To the left the typical hierarchy of contemporary workstation and computer cluster (nodes) with central processing units (CPUs) with multiple cores and general purpose graphics processing units (GPGPUs), respectively. To the right the trade off to make been memory bandwidth and memory size.}
\label{FigHPCResources}
\end{figure}

Another solution to improve the efficiency of APT data post-processing is to use software parallelization, i.e. strategies and tools from scientific computing for APT. Scientific computing is concerned both with the hardware and with the software to utilize computing hardware most efficiently. Figure \ref{FigHPCResources} substantiates the hardware and the memory hierarchy to master on contemporary computers \cite{Hennessy2012}. The recipe of success when working which such computers has three key ingredients: first, all data have to be placed closest in the memory of the computing core that needs it. Second, each data portion that has already been loaded should be reutilized as frequently as possible. Third, each dataset and its processing should be ideally distributed spatially into work packages that are independently executable and load balanced as best as possible. These ingredients work essentially the same for a laptop, a workstation, or a computer cluster, which is an array of connected workstations. 

There are only a few examples which have started to explore above potential of using scientific computing hardware and programming methods for APT \cite{Seal2008,Seal2014,Lu2018,Katnagallu2018}. Maybe this situation has been caused by our traditionally stronger placed focus on the science within the APT data rather than the methods used to extract it or because most APT users are not frequently trained in software engineering and data science. Nevertheless, we are convinced that with above big data trends the situation is going to change for everyone's benefit. In effect,  it is worthwhile to close the gap by adapting to employ scientific computing for APT research.

Efficiency measures quantify how well a software ticks above technical necessities. Central to this work is the concept of strong scaling, i.e. how much faster a task with fixed costs is executed when using more cores. A program shows ideal strong scaling efficiency only if the elapsed time reduces by the same factor as the core count increases \cite{Amdahl1967}. For APT strong scalability is relevant whenever the execution of a single analysis task should be accelerated. By contrast, weak scaling benchmarks situations in which the task costs and the number of cores are equally upscaled for instance by sweeping trivially parallel through an analysis of the same tasks applied to multiple datasets.

Rather than reporting the speed up, we report the efficiency, i.e. the speed up divided by the core count. The term multithreading covers techniques of how to use the multiple cores of the CPUs on a computing node. One level of the hierarchy higher, the term multi-node covers techniques for distributing a program on and coordinate the communication across multiple nodes. Hybrid parallelism combines this scale, i.e. across the cores of a single node and multiple nodes.


Scientific computing could speed up APT data mining by orders of magnitude as well as alleviate the higher numerical costs of the more complex methods. We acknowledge that APT practitioners feel comfortable with using primarily proprietary software, yet many are open to do so in conjunction with a box of highly performing, community-led tools which represent a versatile approach to tackle many of the challenges discussed above. Scripting options through \matlab{} and \python{} allow for assembling these tools into sophisticated workflows \cite{Janssen2019,Draxl2020} which adhere to the FAIR principles. These tools and workflows can in turn be interfaced with \ivas{} \cite{Reinhard2019,Day2019} if necessary and/or facilitated by the instrument manufacturer.

With this paper, we contribute \paraprobe, our first step to develop an open box of scientific computing tools for analyzing APT data. The paper is organized as follows: first, we detail the key steps of the algorithms and how they were parallelized. Second, we showcase the benefit of the tools, targeting primarily the APT practitioners, with a typical characterization study encountered in experiments for alloy design. Third, we report systematic benchmark results for synthetic specimens in a section which targets software developers.

\newpage
\section{Results and discussion}
\label{ResultsDiscussion}

\subsection{Application to experimental datasets}
\label{ResultsExperiments}
\begin{figure}[!ht]
\centering
	\subfloat[]{\includegraphics[width=1.00\textwidth]{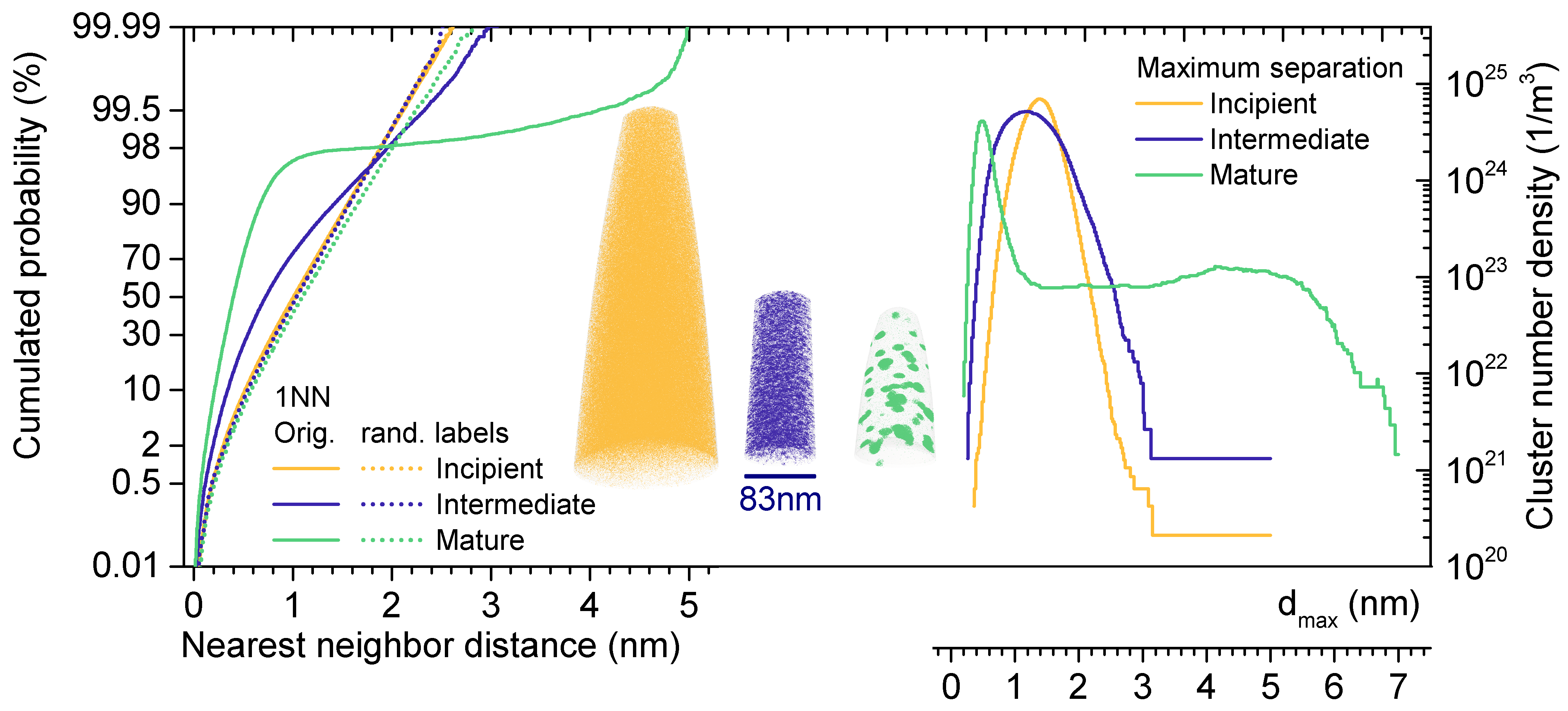}\label{ResExpDescrStats}}
	\quad
	\subfloat[]{\includegraphics[height=0.44\textwidth]{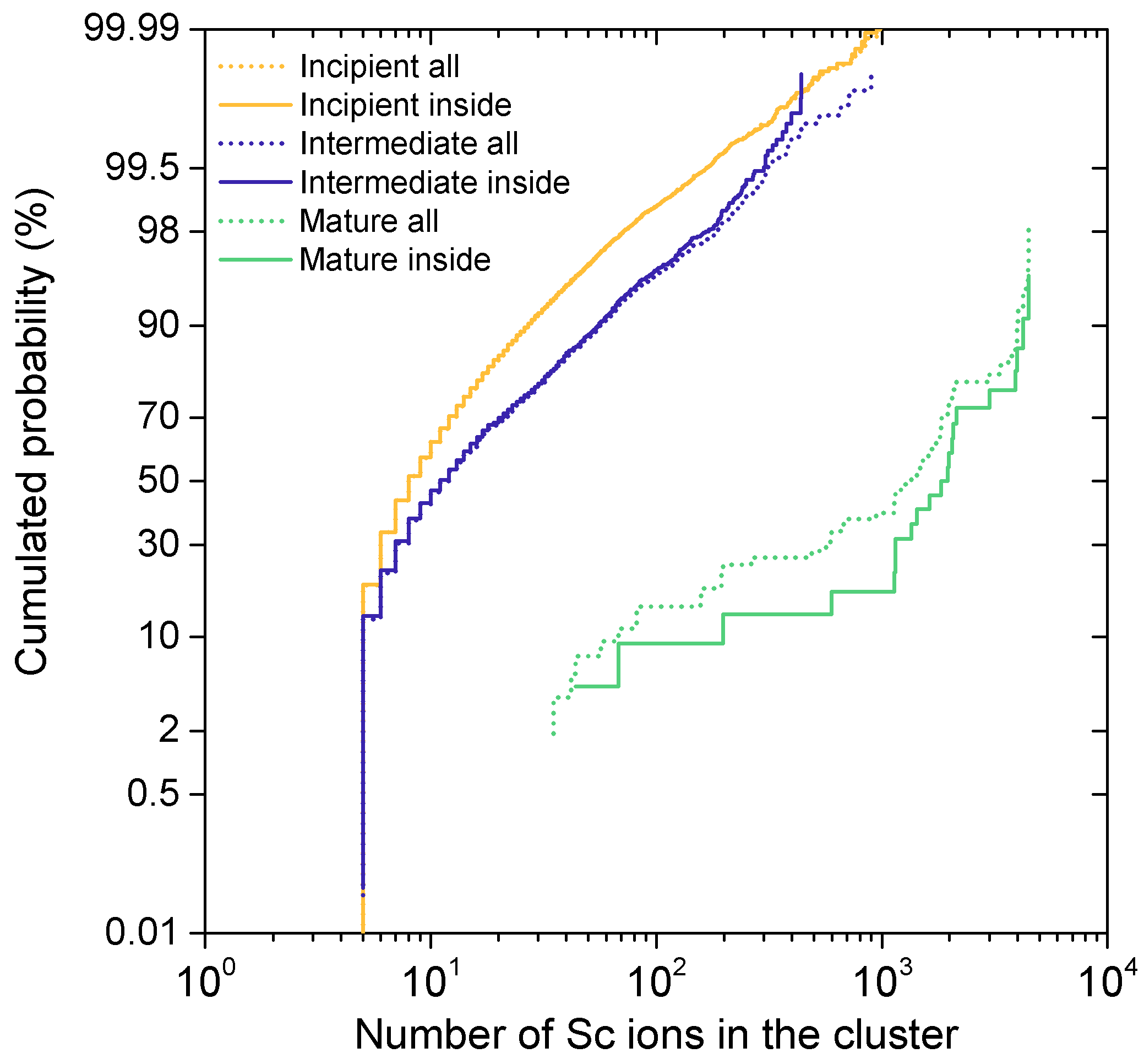}\label{ResExpClustering}}
	\quad
	\subfloat[]{\includegraphics[height=0.45\textwidth]{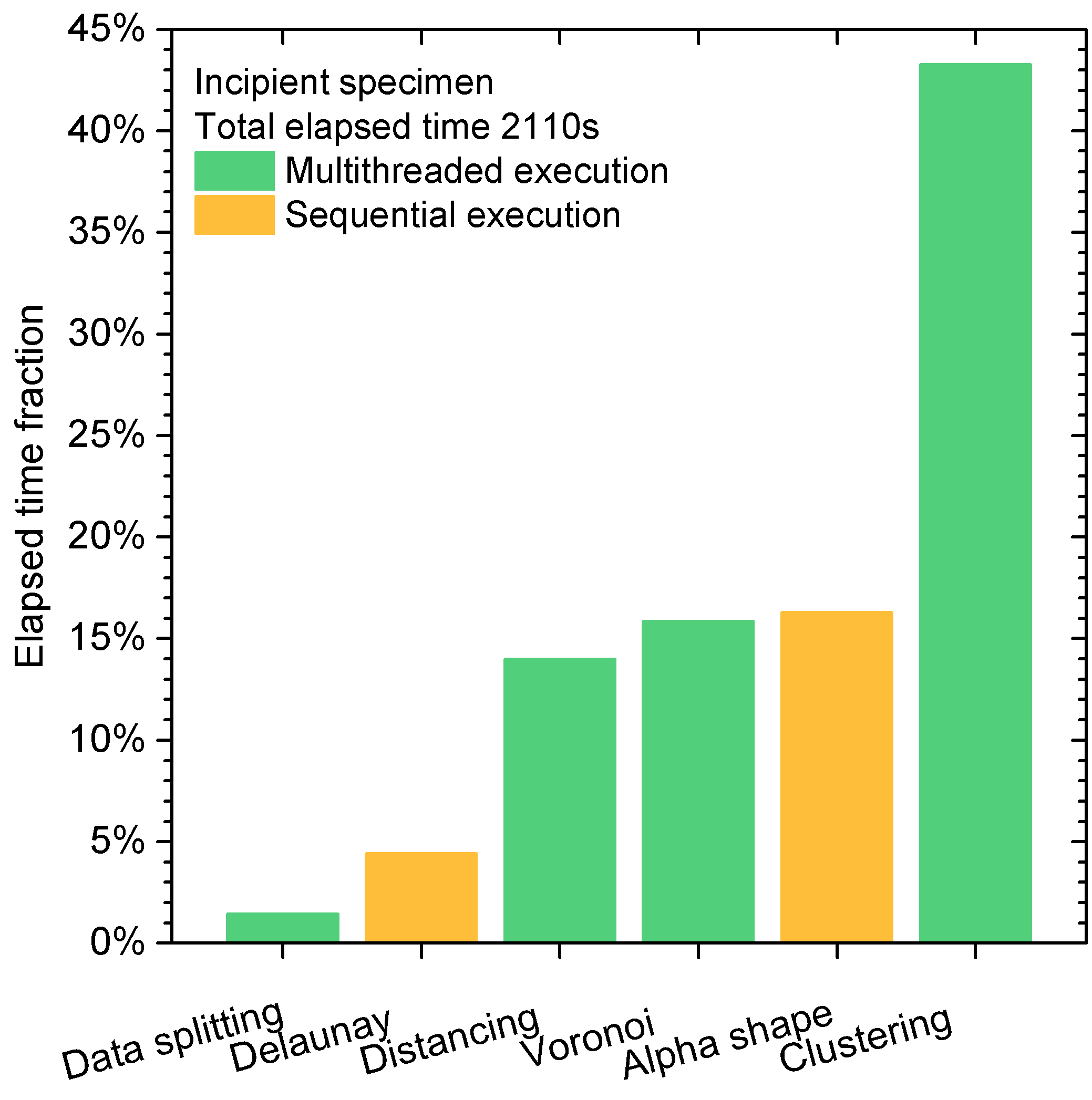}\label{ResExpBenchmark}}
\caption{Sub-figure a) compares the three specimens heads up and rendered at scale. The color-coding distinguishes the specimens and displays their Sc ions. On the left side the 1NN distributions for original (orig.) and randomized (rand.) ion type labels are compared. On the right side the sensitivity of the cluster number density is shown as a function of $d_{max}$. Sub-figure b) compares the distribution of cluster sizes for the specimens and with respect to edge effects for distributions which consider only the interior (solid lines) or all clusters (dashed lines), respectively. Sub-figure c) summarizes the most time consuming parts for the incipient specimen.}
\label{FigResExp}
\end{figure}

The main motivation behind the AM case study is to exemplify how batch processing in combination with parallelization yields additional scientific value faster and with less effort. As a typical example for applying clustering methods, Fig. \ref{FigResExp} summarizes how the maximum separation method performs for the three AM specimens.

The results differ qualitatively and quantitatively. For the datasets from the incipient and intermediate samples, Fig. \ref{ResExpDescrStats} documents that a key assumption relevant to apply the MS method is violated: the individual spatial distribution functions of the ions within clusters do not differ substantially from the distributions of the matrix ions \cite{Stephenson2007,Jaegle2014}. This observation is particularly evident for the 1NN distributions (Fig. \ref{ResExpDescrStats}). Consequently, any interpretation of number densities at virtually all $d_{max}$ values for the incipient and intermediate specimens is inaccurate. This is especially visible for the global maxima of the $d_{max}$ curve. 


An inspection of the cluster size distributions (Fig. \ref{ResExpClustering}) pinpoints the shortcomings of applying the MS method to the incipient and intermediate stage specimens. The distributions show that as many as \SI[mode=text]{25}{\percent} of the identified clusters contain only five Sc ions, i.e. the minimum accepted count $N_{min}$. Again, this is a clear argument against using the MS method for quantifying the early stages of precipitation in those AM specimens.

In contrast, the Sc 1NN distribution for the mature state is bimodal. In this case, the MS method is selective and useful as Fig. \ref{ResExpDescrStats} confirms. It is reasonable to report the plateau value of the curve as the metallurgical relevant number density for two reasons:  first, for this $d_{max}$, most Sc ions contribute to the precipitates rather than to the matrix. Second, for this $d_{max}$, a potential bias in the 1NN distribution due to an accidental fusing of solute Sc ions in the vicinity of the precipitates is lower compared to number densities read at larger $d_{max}$ values. In effect, these results reassure the validity of earlier findings pertaining to the application of the maximum separation clustering method \cite{Stephenson2007,Jaegle2014}. Our work provides additional value by delivering \textit{quasi} unbiased distributions of the cluster size (Fig. \ref{ResExpClustering}). The distributions are \textit{quasi} unbiased because of the capability to detect which clusters were truncated by the dataset edge. 

Given that many APT datasets may contain only a few hundred mature clusters, quantifying this effect is relevant. Figure \ref{ResExpClustering} compares the size distributions for all clusters with the distribution for exclusively those clusters in the interior of the dataset. For the mature state, the shape of both distributions is very similar. Individual quantile values are shifted, though, in particular for the lower half of the curve. We can thus conclude that two contributions --- the truncation of the clusters by the edge and the lower total number of clusters --- in the mature state specimen affect the distributions. By contrast, for the other two specimens the distributions are more similar because of the fact that more precipitates are included in the volume. 

In effect, the AM case study substantiates how \paraprobe{} delivers additional confidence and a detailed quantitative insight on the uncertainty associated with the choice of various parameters. Given that these analyses are instructable with minimal manual interaction, this frees resources of the scientist to discuss rather than wrangle with the data.

The numerical costs of above analyses are summarized in Fig. \ref{ResExpBenchmark}. Executing the parameter sensitivity study on all three specimens took \SI[mode=text]{52}{\minute} when using \SI[mode=text]{36}{} threads. This includes all \SI[mode=text]{241}{} MS clustering runs per specimen, the tessellation, \ashape{} edge computation, and ion-to-edge distancing. Clustering was the most costly task with a total execution time fraction of \SI[mode=text]{43.2}{\percent}. With only \SI[mode=text]{1.0}{\percent} of the total elapsed time, the I/O expenditures were negligible, which is another difference compared to the proprietary tools. The results are encouraging enough to motivate further efforts by the APT community. These should be directed toward similar parameter sensitivity assessments, uncertainty quantification, and making cross-method comparisons with rigorously transparent tools.

Immediate potential for further parallelization is available but has not been tapped in this study. We want to emphasize that all parameter runs were executed sequentially but each run of the MS method executed with multithreading. Alternatively, the parameter runs could be distributed trivially parallel on multiple computing nodes. This would result in hybrid-parallelized execution. The tools could be easily extended to loop in the processing of other clustering methods to allow for fully automatized heads up assessments.

\subsection{Verification and benchmarking with synthetic data}
\label{ResultsSynthetic}
\begin{figure}[!ht]
\centering
	\subfloat{\includegraphics[height=0.44\textwidth]{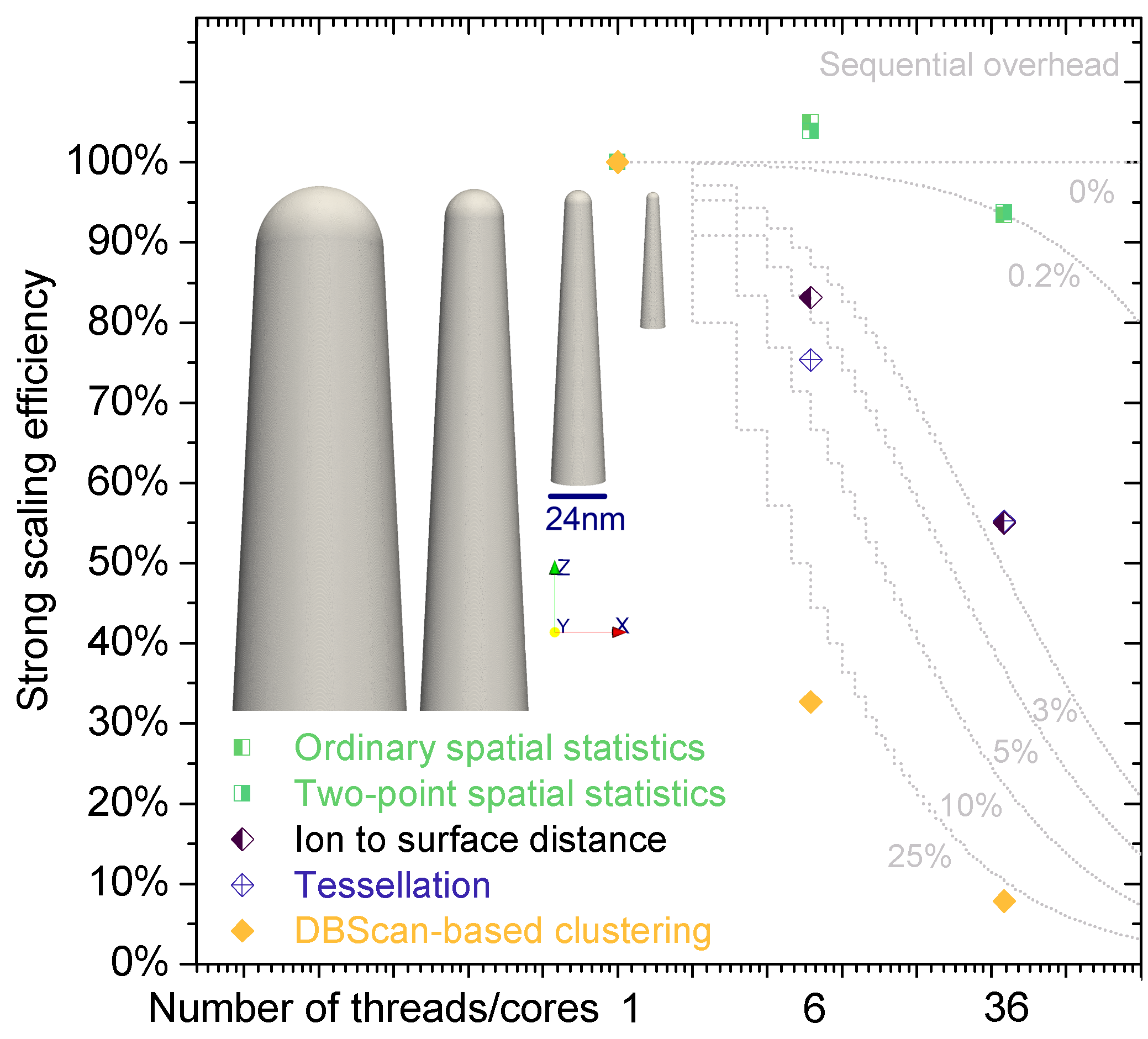}\label{FigResStrgScalEffcyOMP}}
	\quad
	\subfloat{\includegraphics[width=0.45\textwidth]{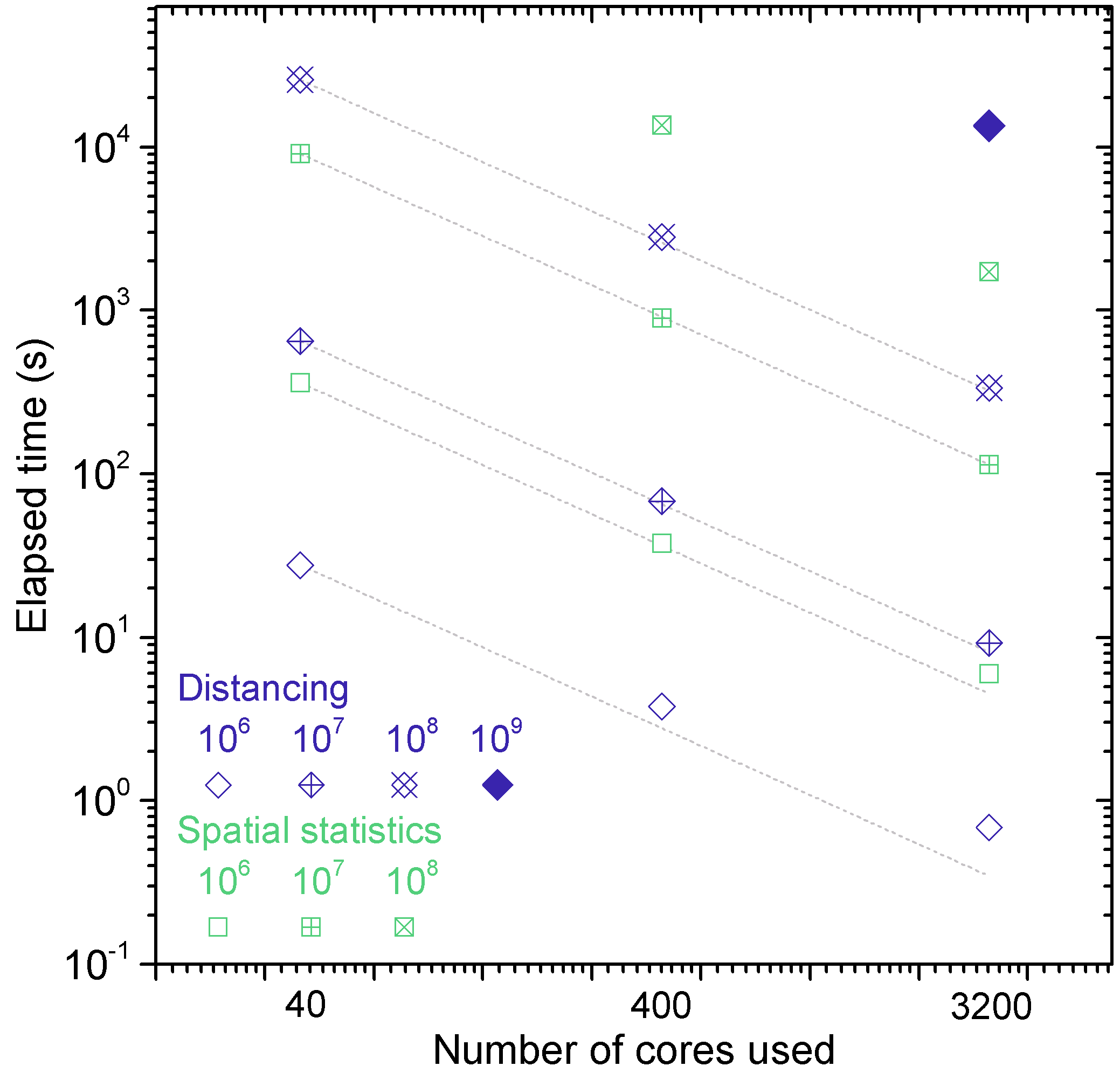}\label{FigResStrgScalEffcyMPIOMP}}
\caption{Strong scaling efficiency for multithreaded a) and elapsed time for hybrid execution. The inset in a) displays the synthetic specimens rendered at scale visualized via the \ashape{} of the edge. For the two leftmost specimens only their upper sections are shown to retain sufficient pixel resolution. The gray curves in a) compare the strong scaling results for different fractions of remaining sequential overhead according to Amdahl's law \cite{Amdahl1967}. In b) the thin dashed lines compare to the theoretical optimum of linearly scaling methods. Runtime differences were within the thickness of the data point symbols.}
\label{FigResStrgScalEffcy}
\end{figure}

\paragraph{Strong scaling efficiency for multithreaded execution}
Figures \ref{FigResStrgScalEffcy} summarize the key results of the performance assessment. Specifically, we report the strong scaling efficiency for multithreaded and hybrid execution. An analysis of the memory consumption is detailed in the supplementary material. To the best of our knowledge, this is the first such assessment of multithreaded APT tools for such a diverse set of analysis tasks. \paraprobe{} shows at least \SI[mode=text]{55}{\percent} strong scaling efficiency when using \SI[mode=text]{36}{} threads for all tasks but clustering. The scaling limitations for clustering are attributable to a sequential overhead as high as \SI[mode=text]{25}{\percent}. One contribution to this overhead is unavoidable because certain steps of the MS algorithm enforce synchronization \cite{Goetz2015}. Another portion of this overhead originates from code sections for the post-processing of the MS results that were computed sequentially. Here, is immediate potential for a further improvement.

For all other tasks also a few percent sequential overhead remain --- despite the fact that we already employed techniques to balance the computational load dynamically. It is this overhead which results in disproportional lower efficiency when using more threads. One key contribution to this overhead is the necessity to accept that the workload per ion, such as during tessellating, typically differs, and thus cannot be perfectly distributed when using many cores.


\paragraph{Strong scaling efficiency for hybrid execution}
Figure \ref{FigResStrgScalEffcyMPIOMP} summarizes the results of combining OpenMP multithreaded with Message Passing Interface (MPI) \cite{Snir1998} process data parallelism. Here, we exemplify an application for distancing the ions to the \ashape{} and processing spatial statistics on an exemplar computing cluster with $80$ nodes with $40$ cores each. The specimens were pristine fully-dense Al single crystals. Their size ranged from \SI{2.0e6}{} to at most \SI{2000e6}{} ions. The distances were computed for $d_{srf} = \SI{10}{\nano\meter}$. We characterized Al-Al spatial statistics for the entire specimens: first, the RDF, second several kNN (with $k = 1, 10, 100$) distributions, and third a complete three-dimensional SDM. The ROI radii were set to $R = \SI{10}{\nano\meter}$ for the RDF and kNN, and to $\SI{2}{\nano\meter}$ for the SDM. Cubic voxel with a volume of ${\SI{0.025}{\nano\meter}}^3$ were employed for the two-point statistics.

The results confirm in all cases that at least one order of magnitude performance gains were achieved in addition to multithreading. The scalability is close to ideal, as it is expected for this moderate number of cores and weakly coupled computation. The more ions each core processes, the more effective the additional parallelization layer becomes. These findings substantiate that using parallelized data mining methods is useful for routine processing of APT tasks. 

\paragraph{\ashape{} computation, ion-to-edge distancing, and tessellating}
In what follows, we summarize the key results for the individual analysis tasks. Primarily, the results address how the elapsed time scales as a function of the specimen volume. For each task we identified specific quantities that are strongly correlated with the computational costs, and therefore the elapsed time.

The time it takes for constructing an \ashape{} from a Delaunay triangulation \cite{Okabe2000} is dictated by  the number of ions which survive the filtering step. The results detail that the proposed filtering method reduces the total number of ions by \SI[mode=text]{88.5}{\percent} for the smallest and up to \SI[mode=text]{98.8}{\percent} for the largest specimen. This enables for the first time to successfully compute \ashapes{} even for APT specimens with two billion ions at all; and doing so in less than an hour without a necessity for \textit{ad hoc} downsampling. The resulting hull contains \SI[mode=text]{6.57e6}{} triangles.

\begin{figure}[!ht]
\centering
	\includegraphics[width=1.00\textwidth]{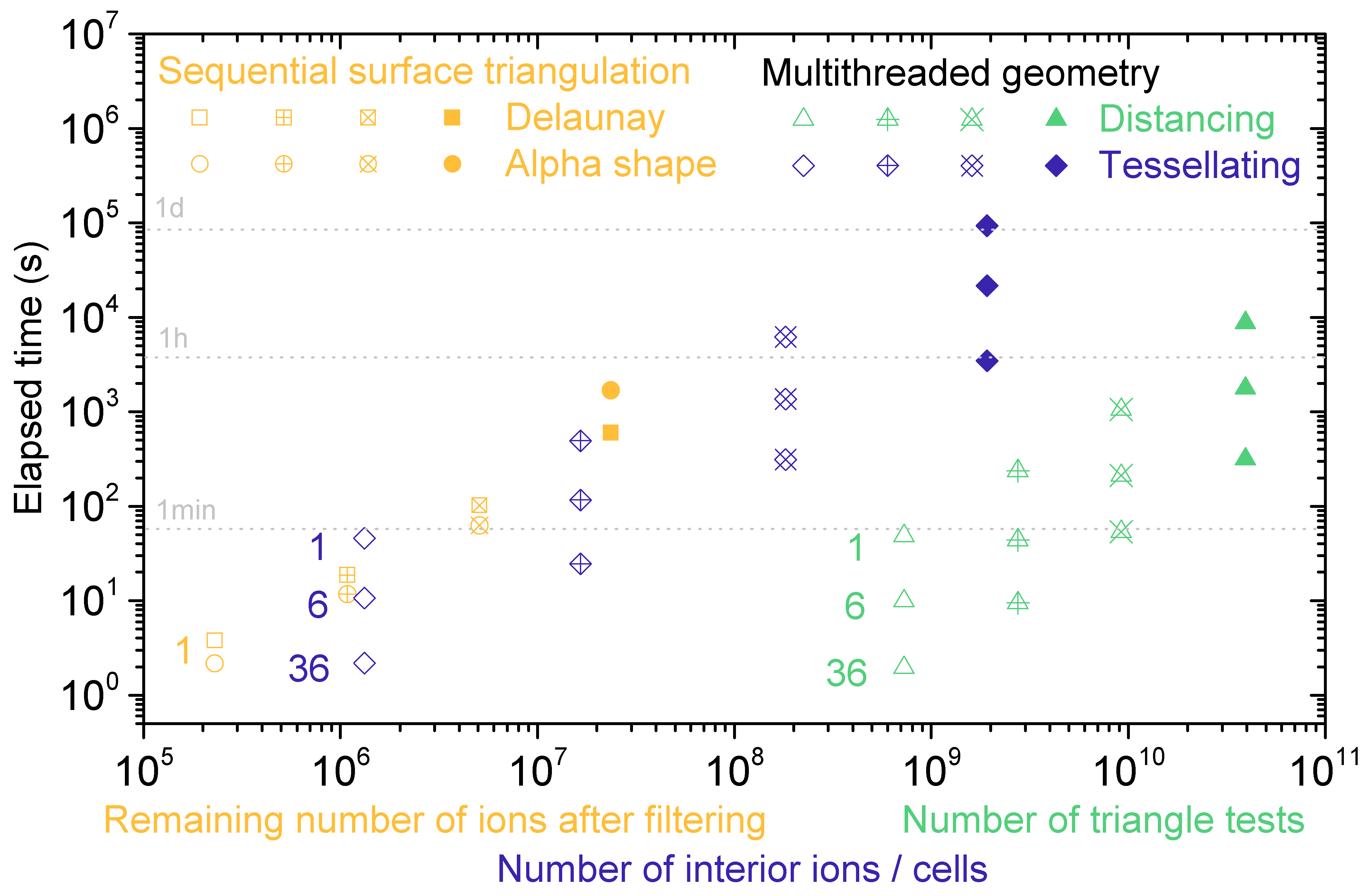}
\caption{Color-coded elapsed time results for the synthetic specimens and different computational geometry tasks as a function of key quantities correlated with increased specimen volume: to the left the triangulation (orange), in the middle the tessellating (blue), to the right the ion-to-edge distancing (green). The data point triplets summarize the results for different number of threads (indicated by the numbers in the diagram). Again runtime differences were within the thickness of the data point symbols.}
\label{FigResCompGeom}
\end{figure}

Figure \ref{FigResCompGeom} documents the performance of the computational geometry tasks, i.e. \ashapes{}, ion-to-edge, and tessellation. The decisive quantity to inspect for distancing is the total number of triangles processed per ion. Using the tree-based triangle filtering strategy reduces the number of individual tests from a million to in many cases as low as \SI[mode=text]{20} triangles per ion on average. Multithreading enables to compute the distances for all specimen sizes up to \SI[mode=text]{27.8}{} times faster than for sequential execution when using $36$ threads.

We benchmark the tessellation tasks for the precipitate-free synthetic specimens. Figure \ref{FigResCompGeom} summarizes the elapsed time. Table \ref{TabSetupBenchCases} lists the specific settings of this analysis. For every ion, i.e. Voronoi cell, the volume, all first-order neighbors, and the shape of the cell was computed. Applying \paraprobe{} with $36$ threads solved this task for the two billion ion specimen in \SI[mode=text]{58}{\minute} -- \SI[mode=text]{27}{} times faster than for sequential execution. To the best of our knowledge, this is the first time that such large tessellations for APT data have been successfully computed. The results are also an independent confirmation and signature of the useful contributions Peterka et al. \cite{Morozov2016} made in their field. This functionality opens up the opportunity to apply tessellation-based algorithms \cite{Felfer2013,Felfer2015b,Felfer2016} now routinely to quantify the volume about each ion or or use the Voronoi cells as building blocks for the reconstruction of microstructural features \cite{Kuehbach2020a}. An extension of our tool for using hybrid parallelism to compute tessellations and perform crystallographic analyses on APT data is in progress. 

\begin{center}
\begin{table}[!htb]
\caption{Multiple spatial and two-point statistics were evaluated for multiple ion type combinations (Al-Al, Sc-Sc, Al-Sc, and Sc-Al) using the batch processing approach.}
\centering
\begin{tabular}{ll}
    \toprule
    \bf{Task}     & \bf{Benchmarking}  \\ \midrule
    \ashape{}       & \ashapeinit{0.5}                                            \\
    Surfacing     & \ionsurf{2.5}                                               \\
    Spatstat      & RDF, kNN ($k = 1, 10, 100$), two-point 1NN ($\Delta r = \SI[mode=math]{0.5}{\nano\meter}$) \\
                    & \spatstatnok{0.0}{0.05}{2.5}{2.5}                              \\
    Tessellation  & \vorotess{1.0}                                              \\
    Clustering    & Sc-Sc, \maxsep{0.10}{0.05}{0.70}{5}                         \\ \bottomrule
\end{tabular}
\label{TabSetupBenchCases}
\end{table}
\end{center}

\begin{figure}[!ht]
\centering
	\includegraphics[width=1.00\textwidth]{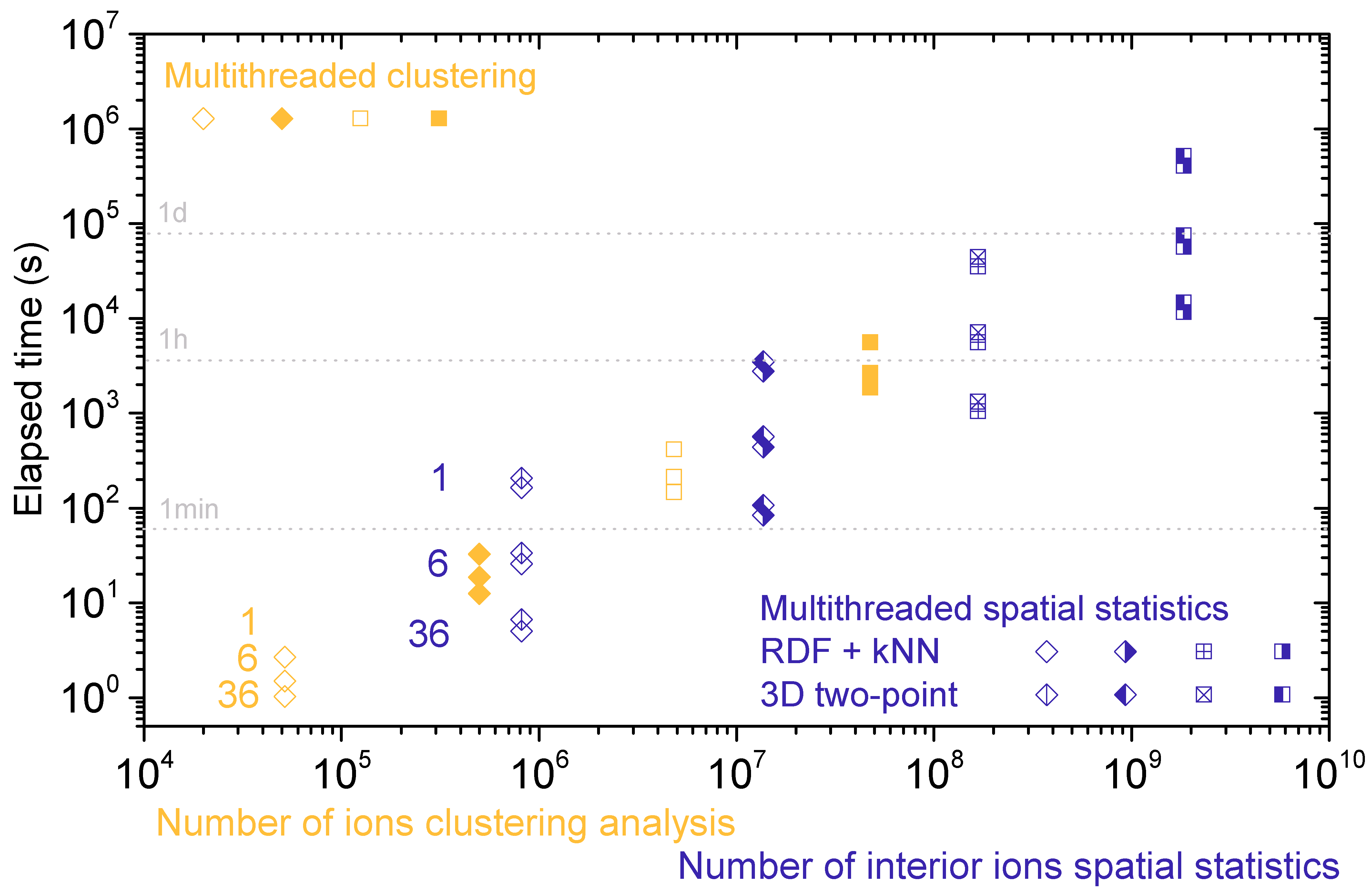}
\caption{Color-coded elapsed time of the multithreaded clustering (orange) and spatial statistics (blue) as a function of the specimen volume.}
\label{FigResStatsClust}
\end{figure}

\paragraph{Spatial statistics and clustering}
In general, \paraprobe{} reutilizes already computed ion-to-edge distances for subsequent analysis tasks. One example is for the quantification of descriptive and two-point spatial statistics. The elapsed time for executing these tasks and performing the MS clustering method on the Sc ions within the synthetic \althreesc{} precipitate specimens (Tab. \ref{TabSetupBenchCases}) is summarized in Fig. \ref{FigResStatsClust}. The diagram distinguishes the results for each task via the color coding.

The collection of orange data points documents the successful analysis of the MS method. Even for the two billion ion tip with a total of \SI[mode=text]{47.6e6}{} Sc ions this analysis took less than an hour. The practical advantages that were identified in the additive manufacturing case study remain: edge and finite counting effects correction capabilities, multithreading, and rigor quantitative analyses. The collection of blue data points substantiates the success of employing multithreading for reducing the elapsed time for spatial statistics tasks by at most a factor \SI[mode=text]{35.1}{} compared to sequential execution. This performance gain in combination with no restrictions on the radius of the ROIs equips APT practitioners now with a powerful tool to process even the largest datasets performantly.

\newpage
\paragraph{Conclusions}
\label{Summary}
To summarize, this paper delivers \paraprobe{}, a box of open source tools for strong scaling data mining of atom probe tomography experiments. Rather than optimized for a single analysis task, our software is a scientific computing tool which solves a number of tasks which atom probers characterize on a daily basis. Exemplified for specimens with at most two billion ions, this work delivers specific parallelized solutions for solving the following tasks:
\begin{itemize}
	\item Build an \ashape{} to the entire point cloud which does not downsample close to the edge of the point cloud via an original filtering algorithm.
	\item Compute exact ion-to-edge distances with which edge effects for spatial statistics and precipitate size distributions can be practically eliminated.
	\item Compute spatial statistics and perform clustering analyses, exemplified here for the maximum separation method, as high-throughput studies via fully customizable and automatically executable batch processing queues. 
    \item Compute a Voronoi tessellation of the entire specimen to assists quantification of atomic level concentration values, and upscale productivity for composing microstructural objects from the Voronoi cells about the ions.
    \item First of its kind proof-of-concept implementation of how to use \hdmf{} as an open file format for APT data. Thereby, our study shows how to achieve improved I/O speed, use better assistance for scientific visualization, and become prepared for studies which seek to better align with the FAIR data stewardship principles.
\end{itemize}

We document at least \SI[mode=text]{55}{\percent} strong scaling multithreading efficiency when using $36$ OpenMP threads. Only clustering scales poorer because of necessary sequential dependencies. These can take as much as \SI[mode=text]{25}{\percent} of the total sequential execution time. With an additional layer of MPI process data parallelism we were able to compute ion-to-edge distances and key spatial statistics (RDF, kNN, SDMs) approximately three orders of magnitude faster than sequentially on a $3200$ cores computing cluster.

\newpage
\section{Methods}
\label{Methods}
\subsection{Computational aspects}
\begin{figure}[!ht]
\centering
	\includegraphics[width=1.00\textwidth]{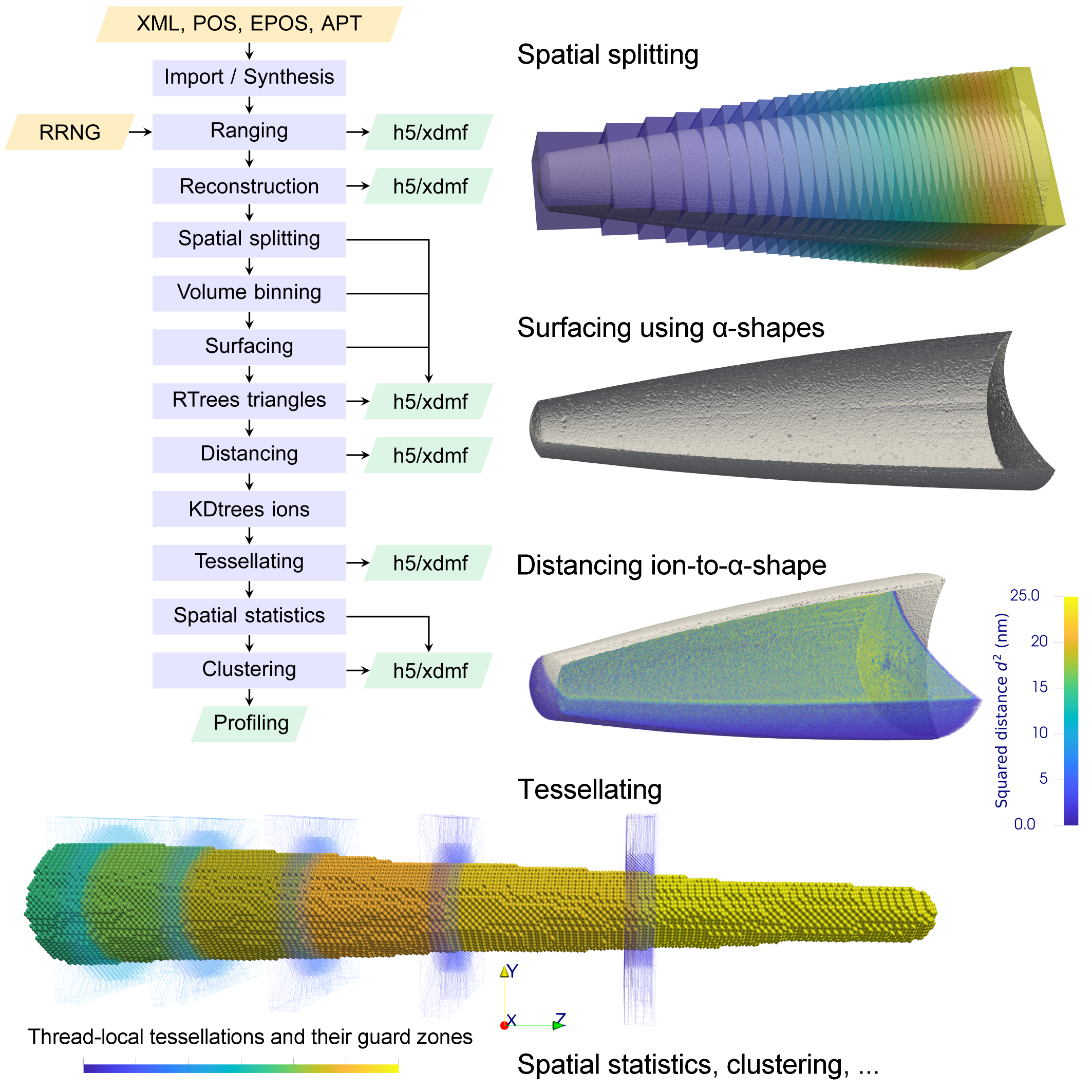}
\caption{\protect\paraprobe{} implements a fully automatized workflow which accepts measured specimens or creates synthetic datasets. Exemplar results of key steps of processing clarify the functioning of the data splitting, ion-to-edge distancing, and tessellation tasks. The guard zones for each tessellated region are highlighted in light-blue wire-frames.}
\label{FigWorkflow}
\end{figure}

\paragraph{Principle design and workflow}
In this section, we detail the types of analyses which \paraprobe{} can perform, the associated workflow (Fig. \ref{FigWorkflow}), and the key techniques for strong scaling performance. To balance between a covering yet concise description, we report all not immediately relevant details of the study in respective paragraphs of the supplementary material.

Instead of a monolithic program, \paraprobe{} is a collection of small parallelized tools. Given that we target workstations and computer clusters, the tools have so far no graphical user interface (GUI). Instead, they are controlled through configuration files which specify the input and the analysis tasks. To assist the users with creating and modifying these configuration files, we developed a web browser-based \python{}/\bokeh{} GUI \cite{Bokeh2019}. An example is provided in the supplementary material.

The raw data of an APT experiment contains a collection of detector hit positions, time-of-flight, and voltage curve data. Typically, single floating point precision suffices. For \cameca{} instruments these results are stored in proprietary container formats (RHIT until \ivas{} v3.6, HITS since \ivas{} v3.8). There is currently no generally working option to parse all content from such files without \ivas{}. Therefore, all analyses with \paraprobe{}, as far as this paper is concerned, were performed in reconstruction space. We rely on \textit{a priori} existing mass ranging information generated for instance with \ivas{} or community tools. A typical run of \paraprobe{} either loads a measurement via standardized formats (POS, EPOS, APT, RNG, and RRNG) \cite{Gault2012b} or builds synthetic specimens. 

\paragraph{Spatial splitting}
Before analyzing, the point cloud is splitted spatially into a stack of non-overlapping cuboidal point cloud regions. We split such along the direction of the longest specimen axis that the regions contain a \textit{quasi} equal number of $N/N_{thr}$ ions \cite{Patwary2016} with $N$ the ion total and $N_{thr}$ threads. Each region administrates its own array of ions. Ions are stored in linear arrays with interleaved position surplus an ion type label. These data demand \SI[mode=text]{16}{\byte} memory per ion. The memory management of the regions was implemented via multithreading, Open Multi-Processing (OpenMP) to be specific \cite{Chandra2001}. Thread pinning and first-touch policies were employed to assure that each thread was executed by one particular CPU core and its arrays stored more frequently in fast accessible memory locations. In effect, \paraprobe{} uses all of the above performance ingredients to accelerate the processing: a combination of data splitting, multithreading, and hardware-topology-aware data placement for improved spatial and temporal locality \cite{Hennessy2012}. 

\paragraph{Quantification of the edge of the point cloud}
Every accurate analysis of spatial quantities for a finite dataset needs a strategy for reducing edge effects. Such can arise when interrogating the long-range neighborhood of ions at the edge of the dataset \cite{Stephenson2007,Haley2009}. One strategy could be to identify the shortest distance of an ion to the edge and use this distance as a criterion to exclude ions from an analysis to avoid bias. The same strategy can be applied to precipitates when they are only partially analyzed, i.e. truncated by the dataset edge. 

One strategy to define the edge is to construct a triangle hull to the point cloud. A variety of methods exist and have been used for APT data: convex hulls \cite{Okabe2000,Haley2009}, \ashapes{} as a generalization of convex hulls \cite{Edelsbrunner1994,Felfer2016}, or \gshapes{} as a generalization of \ashapes{} \cite{Cameron2004}. The benefit of $\alpha$- and \gshapes{} over convex hulls is that they can account for concave sections of the point cloud. 

However, the key challenge why \ashapes{} have hitherto not rigorously been used, especially not for large APT datasets, is that the algorithms for constructing them have poorer than linear time complexity $\mathcal{O}(N^p)$ in the number of $N$ ions and with $p$ a scalar larger than one. In other words, computing \ashapes{} gets overproportional costly the larger $N$ becomes. One strategy to cope with this is downsampling. However, this results in less accurate distances. Another strategy is analyzing a section of the point cloud only. We propose a different strategy.

The key advantage is that it retains the accuracy of the point cloud close to the edge. The key observation we utilize is that most ions in the interior of the point cloud do not contribute a triangle to an \ashape{}. Consequently, a filtering algorithm is proposed which filters out these interior ions. With this strategy, the algorithm needs to compute only the relevant ions of which there are at least two orders of magnitude fewer than interior ions, which is especially effective for the multi-hundred million ion specimens. The details of the filtering algorithm are described in the supplementary material. The subsequent \ashape{} construction has two steps: first, the computation of a Delaunay triangulation of the filtered ion point cloud \cite{Okabe2000,CGAL2018}. Second, the triangulation of \ashapes{} for specific $\alpha$ values. \paraprobe{} executes both steps sequentially.


\paragraph{Ion-to-edge distancing}
The triangulated representation of the edge enables the computation of ion-to-edge distances. Approximate and exact analytical methods can be used. \paraprobe{} computes distances $d$ analytically. The distances can then be evaluated against a threshold distance $d_{srf}$ to eliminate bias in subsequent analysis tasks. The key challenge when computing exact distances is that potentially a large number of ion-to-triangle tests have to be evaluated. Therefore, \paraprobe{} implements a multi-step filtering algorithm which reduces the number of ion-to-triangle tests. The details are reported in the supplementary material. For each ion, first a coarse distance is evaluated. Second, this value is used to identify a smaller set of candidate triangles via an R-tree \cite{Brinkhoff1996,Balasubramanian2012} of the \ashape{}. Details to the implementation of the data structures are given in the supplementary material. All distancing works multithreaded. Ions are machined off region after region. For each region all threads process ions via a dynamic scheduling approach.


\paragraph{Descriptive, two-point spatial statistics, and clustering}
Spatial statistics \cite{Cressie1991} characterize the ions' spatial environment. Applied for APT data, a variety of probability density functions and their distributions are commonly used. Examples are kNN, i.e. the distribution of distances between ions of a certain type and their individual $k^{th}$-nearest neighbor (of a certain type); or the radial distribution function (RDF) \cite{Sudbrack2006,Philippe2010c}, which offers a parameter-free approach for measuring clustering and matrix compositions. Such RDF-based analyses connect to methods for small-angle X-ray scattering \cite{Geuser2011,Zhao2018,Degeuser2020}.

RDF and kNN represent annularly integrated representatives of the more general, so-called two-point (spatial) statistics \cite{Cecen2018}, i.e. functions which quantify three-dimensional probability mass values which describe how many neighboring ions of a particular type the ions have in a particular direction $\vec{r}$ and radial distance $R$. Serial sections of these functions in the central ion's plane of location are better known by atom probers as spatial distribution maps \cite{Geiser2007}. \paraprobe{} implements all above-mentioned spatial statistics with a customizable binning with rectangular transfer functions and hybrid parallelization.

Clustering algorithms applied to APT data \cite{Marquis2002,Stephenson2007,Degeuser2020} enable the quantification of the number density and the distribution of sizes for precipitates or clusters. For reasons of practicality, the term cluster and precipitate will from now on be used interchangeably. A variety of clustering methods has been reported \cite{Stephenson2007,Gault2012b,Ghamarian2019,Zelenty2017,Gwalani2019}. Especially variants and generalizations of the DBScan \cite{Ester1996} clustering algorithm, such as the maximum separation (MS) \cite{Hyde2000,Jaegle2014}, the core-linkage \cite{Stephenson2007}, or the hierarchical DBScan \cite{Ghamarian2019} method are employed for APT data. Given the importance of DBScan variants, we decided to focus in this work on their potential for parallelization. \paraprobe{} thus executes the OpenMP-parallelized DBScan implementation of G\"otz et al. \cite{Goetz2015}. We modified their code to execute the MS method. In effect, \paraprobe{} executes a batch queue of individually multithreaded DBScan runs to assess the parameter sensitivity. If no ion concentration should be extracted, the MS method requires the calibration of two parameters: the threshold distance $d_{max}$ and the minimum number of ions per cluster $N_{min}$.

\paragraph{User-defined batch processing queues}
Our work advantageously addresses practical aspects of automation when computing above quantities. One such is the characterization of different ion type combinations via automatized tools. We refer to a single combination of central ions and neighbors as a spatial statistics query task. \paraprobe{} enables users to \textit{a priori} formulate a list of combinations of multiple statistics, multiple query task combinations, and multiple ion types. At runtime, this task list is machined off with an internal batch queue processor whose details are explained in the supplementary material.

\paragraph{Efficient spatial indices for querying ions}
A spatial index is an auxiliary data structure whose purpose is to speed up the spatial queries for geometrical primitives. These primitives can be points or triangles. The key purpose of the index is to filter out as many irrelevant primitives as possible at lowest numerical costs. This evidently demands for a compromise. Namely, the computing time and memory demands for constructing and storing objects in the index are additional costs; and hence should be as low as possible. However, the better the index is spatially organized the lower are the querying costs. Building a better optimized index demands more computing time and memory. For these reasons, a variety of indices exist. Multiple studies have investigated their (asymptotic) construction and querying costs \cite{Bentley1975,Brinkhoff1996,Balasubramanian2012}, even specific for application on APT data \cite{Lu2018}. \paraprobe{} uses three types of spatial indices: ion queries are accelerated via ion-type-specific KD-trees \cite{Patwary2016} with one tree per dataset region. Triangle queries are accelerated with R-trees \cite{Brinkhoff1996}. Ion queries for clustering and tessellation tasks are accelerated with regular spatial binning. The supplementary material details their specific implementation.


\paragraph{Parallelized volume tessellations}
A tessellation is an overlap-free distributing of a space \cite{Okabe2000} for a given set of points and a mathematical space distributing rule. The rule which defines a Voronoi tessellation, i.e. which assigns each position in space to the individually closest point, yields several useful results for APT data: a defined volume, and thus concentration value per ion \cite{Breen2013,Felfer2015b}, three-dimensional Voronoi cells with a topology which are useful for cluster identification \cite{Felfer2015b}, and cell facets with which microstructural features \cite{Felfer2015a,Felfer2015b,Kuehbach2020a} can be reconstructed.

Despite these benefits, the construction of tessellations for especially the large APT datasets faced so far unsolved challenges because existent computational geometry libraries were used sequentially and out of the box. A more detailed discussion is given in the supplementary material. 

\paraprobe{} breaks with this concept and translates instead an alternative solution for tessellating large datasets from the cosmology into the APT community. The key idea is to split the tessellation task first into multiple smaller tessellations. In a second step, these are fused at the edges. Following this idea of Peterka and coworkers \cite{Morozov2016}, \paraprobe{} splits the tessellation of the entire point cloud into as many tessellations as there are regions. Now these regions are independent and thus processable via multithreading. For this purpose we implemented a multithreaded wrapper around the \voroxx{} library. Each thread processes one region.

Guard zones were attached on either side of the region and exact partial copies of the point clouds from the adjoining regions copied in to ensure that also the cells at the region edge are computed with a correct individual shape. Figure \ref{FigWorkflow} shows an example of these guard zones (light-blue wire-frames) and the resulting tessellation for six threads. Additional details to the implementation, specifically the handling of the cells at the dataset edge are detailed in the supplementary material. 

\paragraph{HDF5/XDMF files for storing the results}
File formats with open specifications offer a transparent way to store APT data. This aligns with the aims of the FAIR data stewardship principles \cite{Wilkinson2016,Draxl2020} as well as with recent activities of the International Field Emission Society Technical Committee. We are convinced that scientific computing file formats like the Hierarchical Data Format (HDF5) \cite{Prabhat2014} offer a more performant tool than the traditional formats and I/O strategies of many APT practitioners. Thus, it was an additional aim of this work to explore the practicality of HDF5 for storing APT data. 

HDF5 is a binary container format. An HDF5 file can store the analyzed data and additional descriptive information, i.e. the meta data. This makes it a promising FAIR-aligned tool for the transport and storage of APT metadata. Compared to the recently proposed and also open APT file format \cite{Reinhard2019}, HDF5 has further advantages: the source code is open, the library offers in-place compression functionality, and is optimized for both sequential and distributed memory parallel I/O. Interfaces for \cxx{} and \fortran{} exist and are particularly intuitive to use for \python{} \cite{Collette2013}, and \matlab{} helping scientists to organize their data.

For these reasons, \paraprobe{} test for the first time to store all output in HDF5 container files. Collecting the results of the batch queue processing in a single rather than a collection of files is faster because it reduces file management costs. To assist the user in visualizing results, \paraprobe{} generates supplementary XDMF text files. Details to these files and visualization strategies are listed in the supplementary material.

\paragraph{Implementation}
We implemented \paraprobe{} as a collection of \cxx{} tools. Analyses were executed on two computers: The multithreaded runs were processed with an in-house $36$ core Linux workstation. The hybrid runs were processed on TALOS, a $80$ node Linux computing cluster with $40$ cores per node. All machines were used exclusively, threads were pinned and placed machine-topology-aware. Details are summarized in the supplementary material.

\subsection{Case studies}
\label{DefCaseStudies}
\paragraph{Quantification of clustering in additively-manufactured (AM) alloys}
We analyzed specimens from additively manufactured samples of a research project on characterizing the effects of intrinsic reheating on the precipitate population during AM operations. The specimens contained clusters and precipitates in different growth states. Above states are referred to as the incipient, the intermediate, and the mature state, respectively (Fig. \ref{FigResExp}).

The incipient and intermediate samples were produced via Directed Energy Deposition (DED) \cite{ISOASTM2015} from an \alscsi{0.49}{0.45} \itswtpercent{} alloy in the as-produced state. Clusters in the DED sample formed in response to the intrinsic reheating of the deposited layers during AM. The specimens were taken from the bottom and the top parts of the sample, yielding the incipient and the intermediate states, respectively. The mature state originates from an \alscsi{0.44}{0.02} \itswtpercent{} alloy which was processed via Laser Powder Bed Fusion (L-PBF) \cite{ISOASTM2015}. After building by AM, the mature state sample was heat-treated at \SI[mode=text]{350}{\celsius} for \SI[mode=text]{10}{\hour}. In response to this aging treatment, the precipitates grew to an average diameter of \SI[mode=text]{20}{\nano\meter} approximately. The specimens were prepared with a lift-out procedure after Xenon ion milling with an FEI Helios PFIB dual beam focused ion beam scanning electron microscope \cite{Zhao2018b}. 

With the desire to use the APT results for injection into mean-field kinetic models \cite{Robson2004}, we characterized the distribution of the precipitate sizes with the maximum separation method \cite{Hyde2000,Jaegle2014}. Facing clusters of different diameter and non-negligible Sc and Si content in solid solution motivated to run a parameter sensitivity study with 241 different $d_{max}$ values per dataset. Subsequent to each parameter run, the cluster labels were used to identify all Sc ions that remained unclustered and from which the 1NN distribution of Sc-Sc ion pairs was computed. Table \ref{TabSetupExpCases} summarizes the analysis tasks. They were executed as a single batch queue per dataset. The number of Sc ions in the specimens ranged from \SIrange{8.64e4}{67.4e4}.

\begin{center}
\begin{table}[!htb]
\caption{Spatial statistics were computed for original (org) and randomized (rnd) ion type labels. Post-MS means after the maximum separation method. Volume binning used cubic bins with $d_{bin}$ edge length. Ions with a distance of at least $d_{srf}$ and $d_{ero}$, respectively to the \ashape{} were considered for spatial statistics and tessellation, respectively.}
\centering
\begin{tabular}{ll}
	\toprule
    \bf{Task}          			& \bf{AM case study}                                        \\ \midrule
    \ashape{}                     & \ashapeinit{0.5}                                          \\
    Surfacing                   & \ionsurf{2.0}                                           \\
    kNN                         & Sc-Sc, org/rnd, \spatstat{1}{0.0}{0.001}{5.0}{2.0}            \\
    Tessellation                & \vorotess{1.0}                                            \\
    Clustering                  & Sc-Sc, \maxsep{0.20}{0.02}{5.00}{5}                       \\ 
	Post-MS kNN        			& Sc-Sc, \spatstatnor{1}{0.0}{0.001}{5.0}                 \\ \bottomrule
\end{tabular}
\label{TabSetupExpCases}
\end{table}
\end{center}

\paragraph{Benchmarking all functionality with synthetic datasets}
Reliable verification and benchmarks of a program call for ground truth data. Therefore, we created two sets of four synthetic specimens (Fig. \ref{FigResStrgScalEffcyOMP}). Each was built as a conical frustum with a spherical cap on top \cite{Kuehbach2019a}. Using a fixed shape, the specimen volume was scaled to create datasets with $\SI[mode=math]{2e6}{}$, $\SI[mode=math]{20e6}{}$, $\SI[mode=math]{200e6}{}$, and $\SI[mode=math]{2000e6}{}$ ions, respectively. Pure Al single crystals were constructed with a lattice constant of $\SI[mode=math]{0.404}{\nano\meter}$ without positional noise and fully occupied lattices.

Spherical \althreesc{} precipitates were spatially randomly dispersed into the second synthetic specimen quartet. A lattice constant of $\SI[mode=math]{0.410}{\nano\meter}$ and a precipitate radius of  \SI[mode=text]{2}{\nano\meter} was assumed for replacing Al ions of the matrix by \althreesc{} lattices. The total number of precipitates was scaled linearly in proportion to the specimen volume to contain a total of $\SI[mode=math]{2.69e2}{}$, $\SI[mode=math]{2.52e3}{}$, $\SI[mode=math]{2.41e4}{}$, and $\SI[mode=math]{2.36e5}{}$ precipitates, respectively. Table \ref{TabSetupBenchCases} summarizes all analyzed tasks on these synthetic specimens.

\section{Code availability}
\sloppy
The source code of \paraprobe{} and a documentation is maintained online:
\begin{itemize}
    \item \href{http://gitlab.mpcdf.mpg.de/mpie-aptfim-toolbox/paraprobe}{http://gitlab.mpcdf.mpg.de/mpie-aptfim-toolbox/paraprobe}
    \item \href{http://paraprobe-toolbox.readthedocs.io}{http://paraprobe-toolbox.readthedocs.io}
\end{itemize}
The repository contains also CPU- and GPU-parallelized tools for atom probe crystallography that will be assessed in a future study.
\fussy

\section{Data availability}
The entire repository of compressed data for all benchmarks occupy several terabytes. Therefore, we have splitted the repository, for practical reasons into the configuration, settings, input POS, and essential results files. These files are available online \cite{Kuehbach2020b}. The other results are available from the authors upon serious request.

\section{Acknowledgements}
Computer administration advice from Berthold Becksch\"afer and Achim Kuhl is appreciated. The work catalyzed from scientific discussions with Andrew Breen, Baptiste Gault, Leigh Stephenson, and Franz Roters on how to professionalize tools for APT.

\section{Competing Interests}
The authors declare to have no competing interests.

\section{Contributions}
MK leads the \paraprobe{} project. He designed the tools, implemented the code, performed the analyses, and wrote the manuscript. PB contributed the experiments and corresponding manuscript section. MHC implemented a proof-of-concept GUI for the tools, which MK developed further into the current GUI. EJ and BG contributed through continuous scientific advice, manuscript writing suggestions, and proof reading.

\section{Funding}
MK gratefully acknowledges the funding and computing time grants through BiGmax, the Max-Planck-Society's Research Network on Big-Data-Driven Materials Science and the funding from the German Research Foundation through project RO 2342/8-1.

\newpage
\section{Supplementary material}
\begin{figure}[!ht]
\centering
    \includegraphics[height=0.6\textwidth]{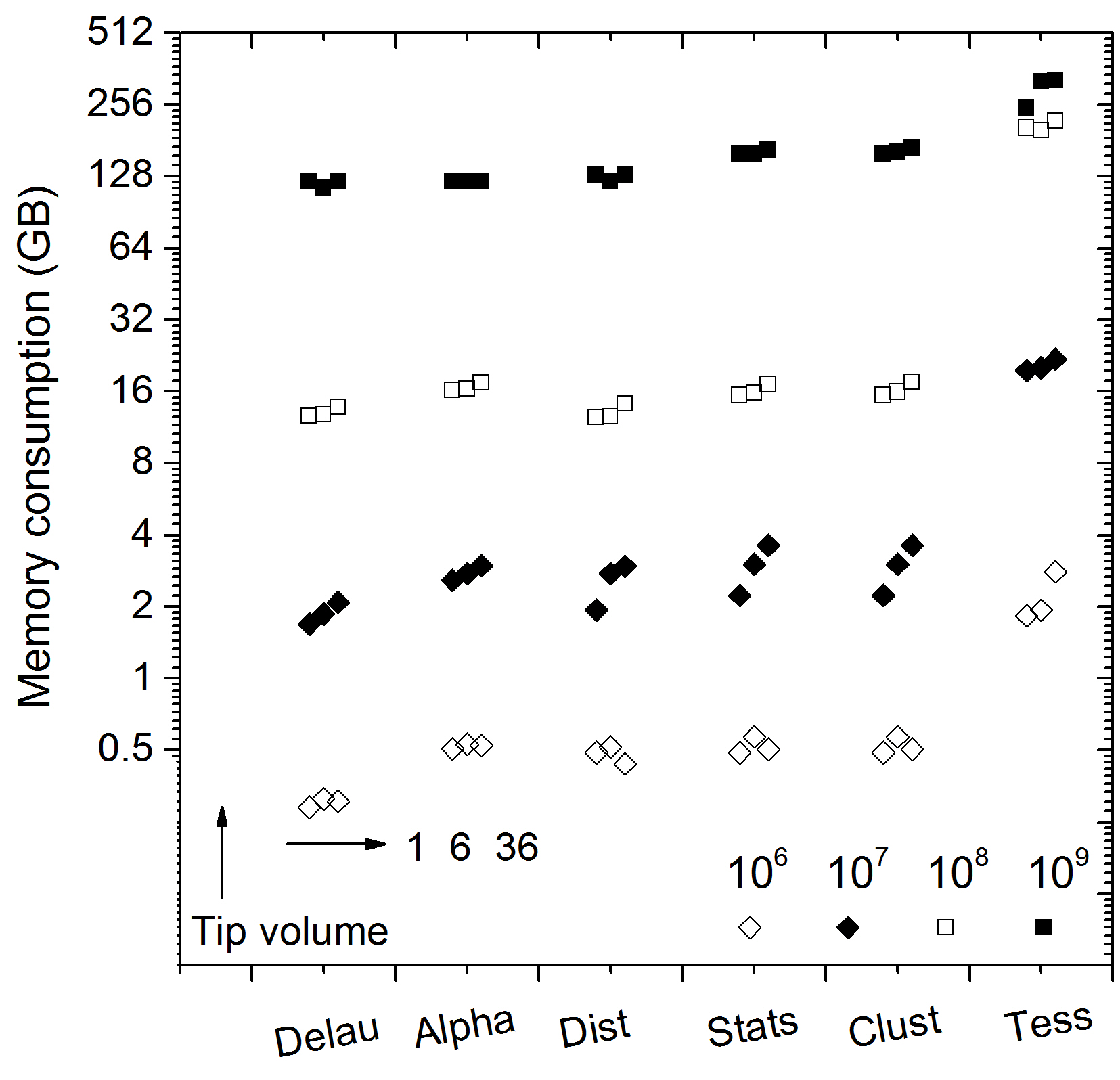}
\caption{Inspecting the peak memory consumption summarizes that, except for potentially the largest setup, all examples were processable with typical workstations. The number triplet 1 6 36 details the number of threads. The numbers in the legend detail the size of the (synthetic) specimens in atoms.}
\label{ResStrgScalEffcyMem}
\end{figure}

\paragraph{Memory consumption for the synthetic benchmarks}
Figure \ref{ResStrgScalEffcyMem} summarizes the peak memory consumption of \paraprobe{}. 
We report the per-page-allocated-memory, which accounts for memory fragmentation, temporary data copies, and all amortized storage of internal program states. These data were parse on-the-fly via the ``/proc/self/stat'' kernel files for the ``SYSTEMSPECIFIC\_POSIX\_PAGESIZE'' virtual and resident memory consumption. The point triplets in Fig. \ref{ResStrgScalEffcyMem} report for successively higher thread counts from left to right. \paraprobe{} consumes at most \SI[mode=text]{270}{\byte} per ion approximately. As such, typical desktop computers with e.g. \SI[mode=text]{32}{\giga\byte}, as it is common in 2020, are sufficient to process all but the largest datasets. Analyzing a dataset containing for instance two billion ions calls for using a larger workstation or a computing node of a computer cluster. It is realistic to optimize our implementation to cut the memory consumption by at least a factor of two in the future by earlier and more rigorous release of temporary data.

Out of all investigated tasks the computation of a complete tessellation has the highest demands of main memory. The reason is that the vertices and the facet polygons for the Voronoi cells were computed for each ion and cached prior I/O. Therefore, at some point the entire tessellation accumulates in main memory. An optimization is possible here in the future with adding intermediate I/O. For instance, the \hdmf{} file for the \SI{200e6} ion specimen stores the shape and topology of \SI[mode=text]{249e6}{} Voronoi cells. This takes up \SI[mode=text]{190}{\giga\byte}.

\section{Details on the computational methods}
\paragraph{Effectiveness of the spatial splitting for heterogeneous datasets}
Preliminary tests showed that datasets with variations in point density and a heterogeneous spatial distribution of different ion types, due to concentration gradients or second phase percipitates, show different efficiency of the processing when using the spatial splitting strategy. No spatial splitting strategy is expected to work equally efficiently for all possible combinations and settings of analysis tasks for heterogeneous APT specimens. Therefore, \paraprobe{} splits the ions of each region further and stores eventually one array specific for each ion type and region.

\paragraph{Ion filtering algorithm}
The filtering algorithm is a labeling algorithm with five steps. All steps work multithreaded. First, we compute a volume binning of the ion point cloud using a rectangular transfer function \cite{Hellman2000b}. Second, the bin aggregate is encased by a one-bin-thick shell of empty bins. Third, the  bins are pre-labeled as either occupied, if they contain at least one ion, or as empty. Fourth, the pre-labels are exchanged by identifying which bins are empty and lay outside the point cloud, which bins are not empty and sample the edge, and which bins are interior, regardless whether they are empty or not. This step is a combined noise cleaning and edge identification step. The cleaning is necessary because if every non-empty bin in contact with an empty bin would be defined as a bin at the edge, also every spurious volume inside the specimen would define a possible edge to the point cloud. In order to avoid this robustly, we perform a percolation analysis on the bin occupancy image in the fourth step. A sequential Hoshen Kopelman (HK) \cite{Hoshen1976} cluster labeling algorithm was used for this task. Fifth, the cluster labels are evaluated against the label of the encasing layer to identify which empty bins are outside, i.e. exterior bins. By virtue of construction, all bins in the encasing layer and their connected bins form a cluster. 

Inverting now the HK result yields which bins compose the interior and which bins the edge of the point cloud. Specifically, the edge bins were identified by scanning their Moore neighborhood \cite{Moore1956} for possible exterior bins. In effect, it remains to pass all ions inside the edge bins to the \ashape{} construction.

\paragraph{Disadvantage of using \ashapes{} for edge computation}
Using \ashapes{} as a representation of the dataset edge has in general, one disadvantage: \ashapes{} are not necessarily closed manifolds, i.e. not necessarily watertight. Indeed, in cases where the density of ions in the point cloud is locally very low, it is occasionally possible that only the \ashapes{} with very small $\alpha$-values, i.e. the \textit{de facto} convex hulls, yield a watertight triangle hull. In other words, it is possible that the triangle hull of an \ashape{} contains holes. Therefore, it might not be possible in every case to compute the volume of the reconstruction based on the triangle hull polyhedron. However, computational geometry methods such as combinatorial mesh repair methods can be used to stitch these holes. Alternatively, one could compute \gshapes{}, which essentially use different $\alpha$-values locally. 

\paragraph{Triangle filtering methods for ion-to-edge distancing}
First, the point cloud is binned with a rectangular transfer function and cubic bins. Second, the labeling algorithm is used to identify bins which form the edge of the point cloud. The user can define the bin width to control this coarse graining. Third, we compute the distance of each bin to the closest edge bin and add a safety distance. Values between one and two times the length of the bin diagonal are practical. As a result, this binning yields a coarse distance mapping. Each bin typically contains a few dozen to a few hundred ions. Second, the coarse distances are evaluated to reduce the number of triangle to be tested during the exact distancing computations. 
With this strategy, the additional costs for the coarse distancing are well overcompensated because of more short-ranged triangle region queries in particular for those ions in the interior of the dataset for which otherwise one would have to take half the base radius of the specimen to come up with a querying radius. In fact, the coarse distance allows to define a local filtering radius how large a region-of-interest (ROI) about each ion needs to be inspected to detect that particular triangle which contributes the shortest distance. Third, we evaluate for all ions within the same coarse bin the same filtering radius. 

\begin{figure}[!ht]
\centering
	\subfloat[a][kNN]{\includegraphics[width=0.3\textwidth]{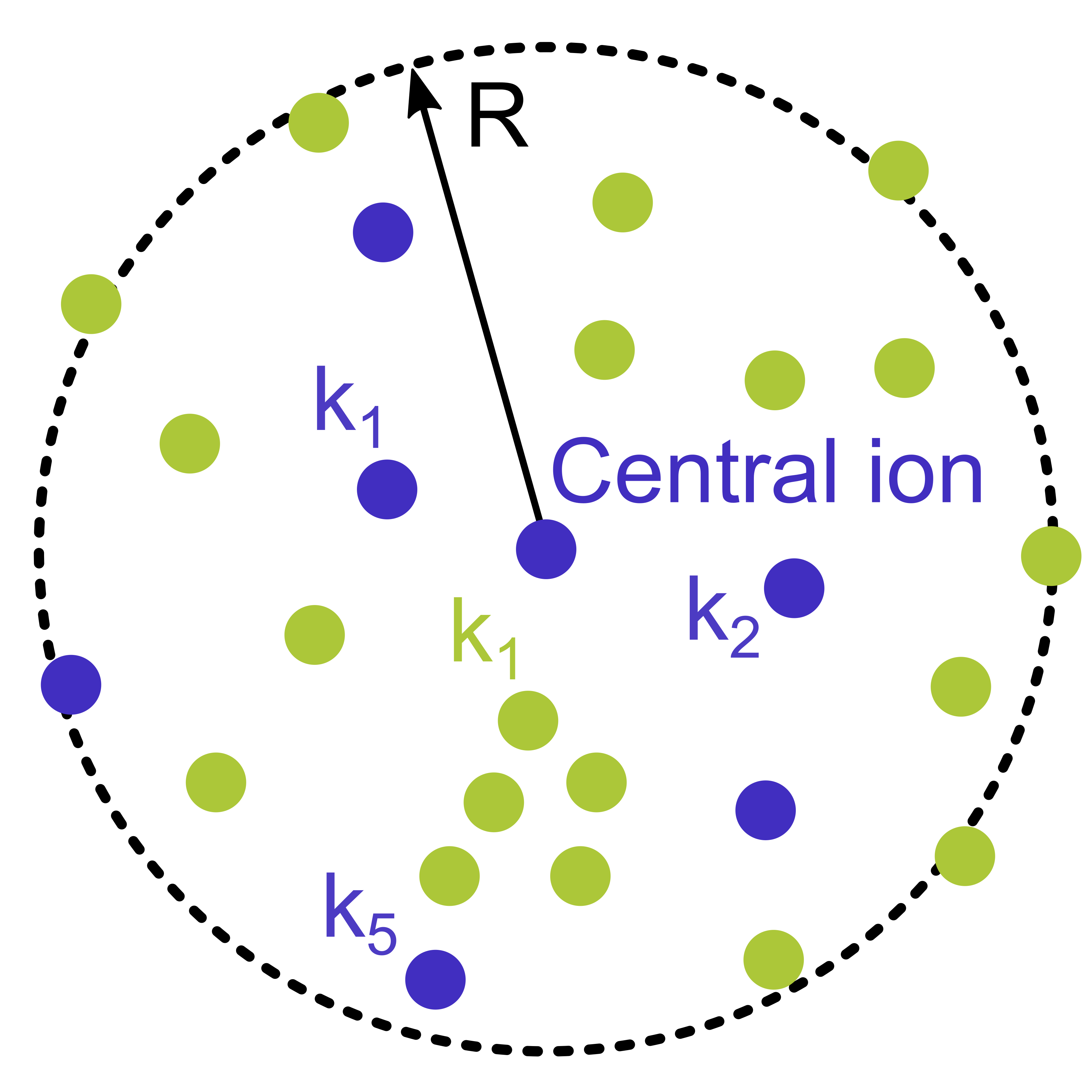}\label{SchematicKNN}}
	\subfloat[b][RDF]{\includegraphics[width=0.3\textwidth]{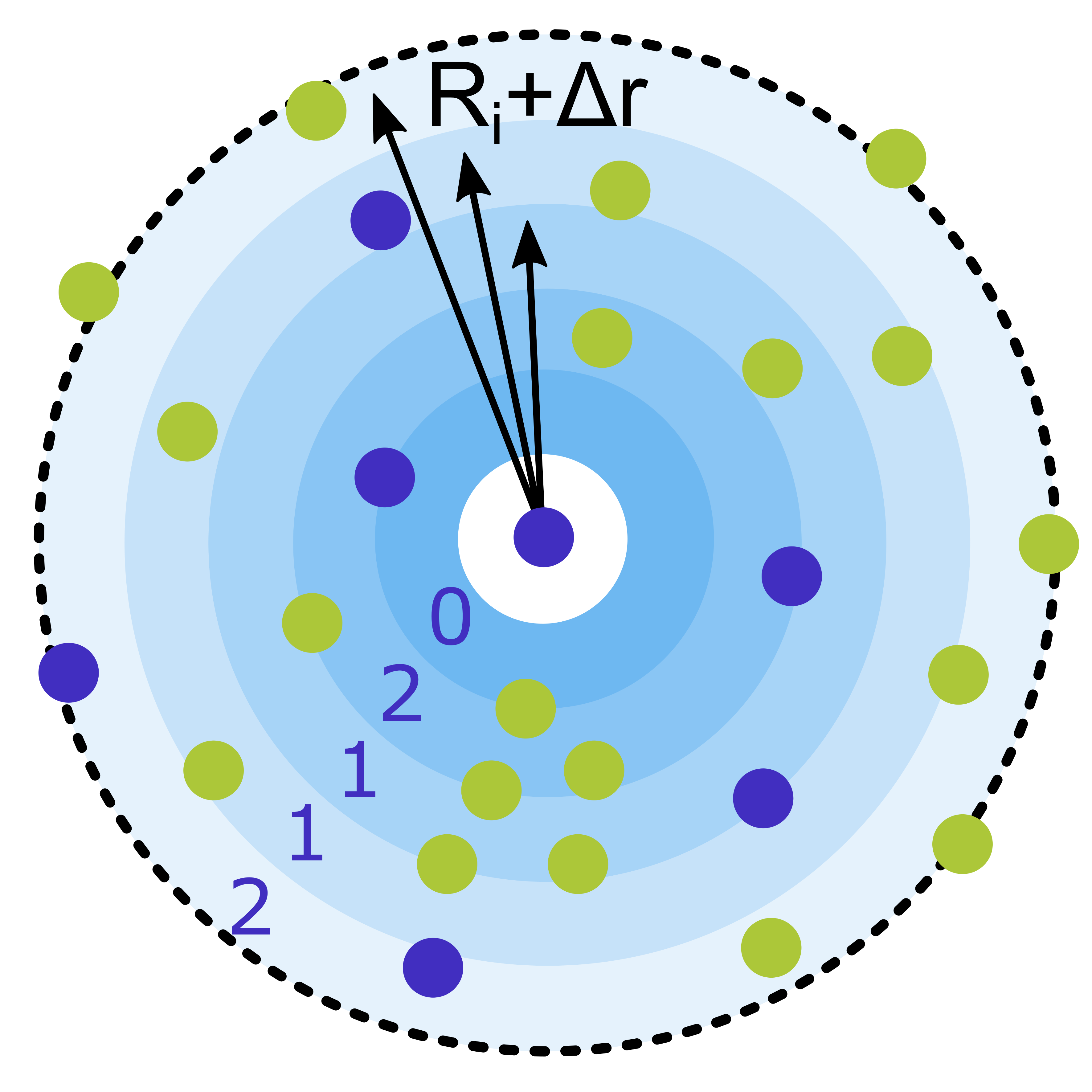}\label{SchematicRDF}}
	\subfloat[c][DBScan]{\includegraphics[width=0.3\textwidth]{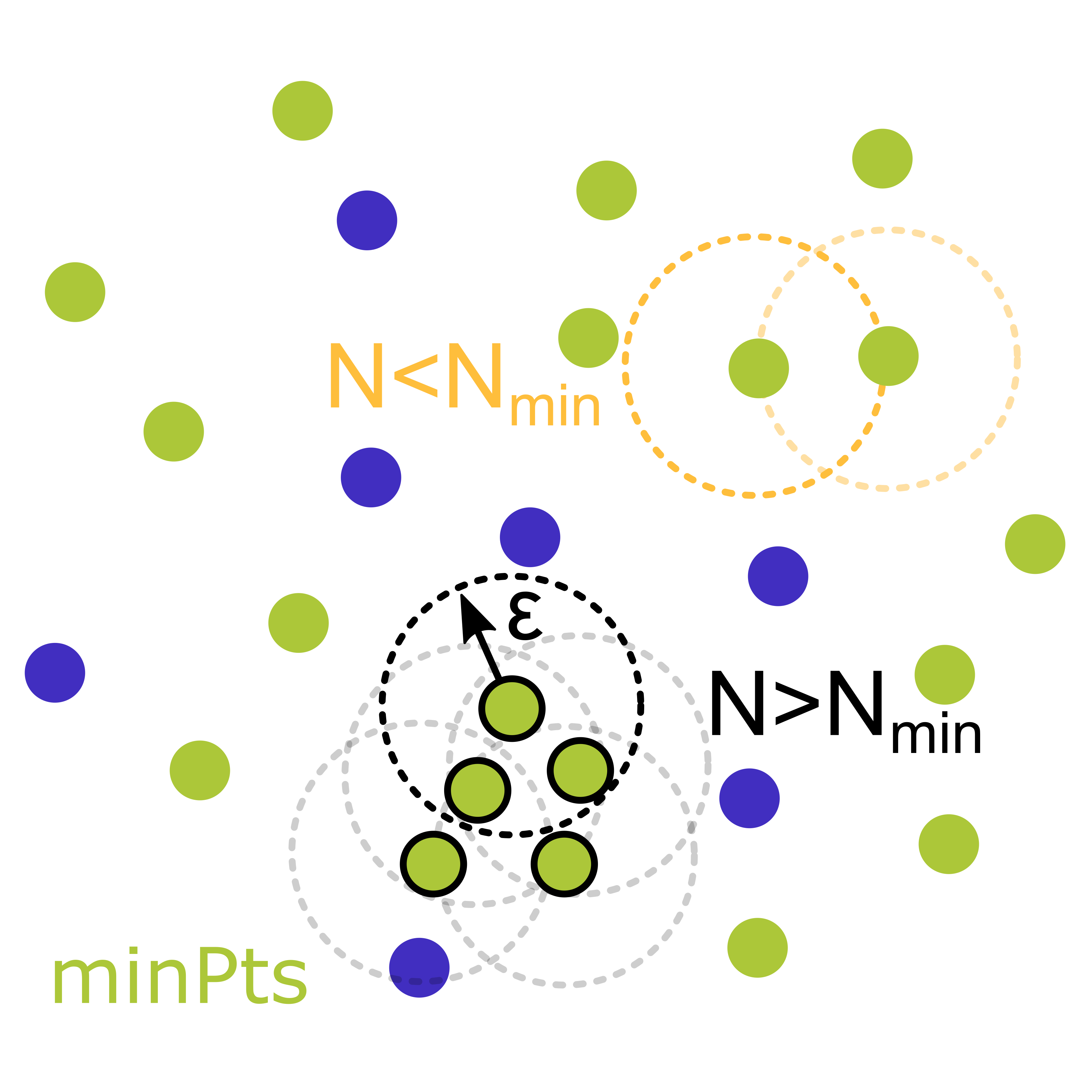}\label{SchematicMAXSEP}}
\caption{Commonly employed spatial statistics are a) $k^{th}$-order nearest neighbor (kNN), especially the case $k=1$, the nearest neighbor (1NN) and b) the radial distribution function (RDF). Clustering algorithms, like the maximum separation (MS) method, group and label ions into objects of potential physical significance. The grouping is based on an analysis which neighbors to an ion lay within an $\epsilon$ distance. Many clustering algorithms set additional constraints when to call such a group of ions a significant cluster: for the MS method, such constraint is the minimum number of ions $minPts$, i.e. $N_{min}$ respectively building the cluster.}
\label{SchematicDescrSpatStatFuns}
\end{figure}

\paragraph{Efficient batch processing of spatial statistics}
Figure \ref{SchematicDescrSpatStatFuns} schematically depicts definitions for the computation of several descriptive spatial statistics functions and the maximum separation (MS) clustering method.

\paraprobe{} solves practical limitations of previous MS implementations and offers improvements which economize parameter sensitivity studies: First, we combine sequential with parallel optimization to gain efficiency. Also we lift the 32-bit address space limitations of a previous MS implementation \cite{Stephenson2007}. Second, the same concept of batch queues was also implemented for clustering. These queues enable the user to instruct automatized parameter sensitivity studies with multiple combinations of MS parameters, as well as multiple combinations of ion types. Third, \paraprobe{} uses its capability to detect the edge of the point cloud to identify which precipitates contain ions in edge bins. This detection enables practitioners to quantify the bias of the precipitate size distribution by either considering all precipitates or excluding the likely truncated ones in contact with the edge of the dataset. Figure 1 in the main paper exemplifies this.

\paragraph{Customized user defined batch processing queues}
Our work advantageously addresses practical aspects of automation when computing above quantities. One such is the characterization of different ion type combinations via automatized tools. We refer to a single combination of central ions and neighbors as a spatial statistics query task. Taking an Al-Sc-Zr specimen as an example, one may wish to compute the $k^{th}$-nearest Sc or Zr neighbors to all Al central ions. The same user may wish to compute, for the same data, further tasks, for instance characterize the $k^{th}$-nearest Sc neighbors of all Zr central ions, and ideally apply these tasks to multiple specimens in one call.

\paraprobe{} enables users to formulate these combinations of multiple statistics, multiple query task combinations, and multiple ion types. The tool implements an internal batch queue processor which machines off a list of querying tasks per ion and dataset. This brings two advantages: first, no repetitive GUI interaction is necessary nor any manual transfers of intermediate results. Both of which are in many cases tedious and error-prone tasks. Second, the \textit{a priori} knowledge of the combinations enables an optimized algorithm which re-evaluates rather than re-queries ions. This reduces the computational costs per ion despite creating higher book-keeping costs for managing multiple results buffers. 

Technically, the statistics are characterized via inspecting the respective region-of-interest (ROI) about all central ions of the requested types. \paraprobe{} identifies neighbors to a central ion within spherical ROIs. The ROIs per region are machined off multithreaded. To compare with a random spatial distribution of the ion types, \paraprobe{} computes by default all spatial statistics for the original and the randomized labels. 

The parallelization concept builds on the observation that each ROI within each region can be processed independently. Consequently, we summarize statistics successively by inspecting the ions of each region of above spatial splitting. For each ROI to an ion, we work through the \textit{a priori} defined list of query tasks. OpenMP multithreading with dynamic scheduling was used to machine off the ROIs per region.

To reduce synchronization across the threads to a minimum, each thread maintains an own set of result buffers. These buffers are summarized to a global set of buffers after each region has been processed. Classical sequential optimization strategies were implemented to reduce the computational costs when querying which ions lay inside which ROI.

\paragraph{Importance of high quality random number generators}
By default, \paraprobe{} computes all spatial statistics for the original and a randomized set of the ion type labels. The labels are randomized with a deterministic pseudo random number generator (PRNG) \cite{Matsumoto1998,Saito2008}. By default, we compute the ROI for both the original ion positions and a configuration with randomized order of the ion type labels \cite{Gault2012b}.
The utilization of a high quality random number generator, like the one above, is important because it reduces bias and inaccuracies when comparing the computed spatial statistics to an analytical equation for random point processes and sample realizations of such point processes.

\begin{figure}[!ht]
\centering
	\subfloat[a][KD-tree]{\includegraphics[width=0.30\textwidth]{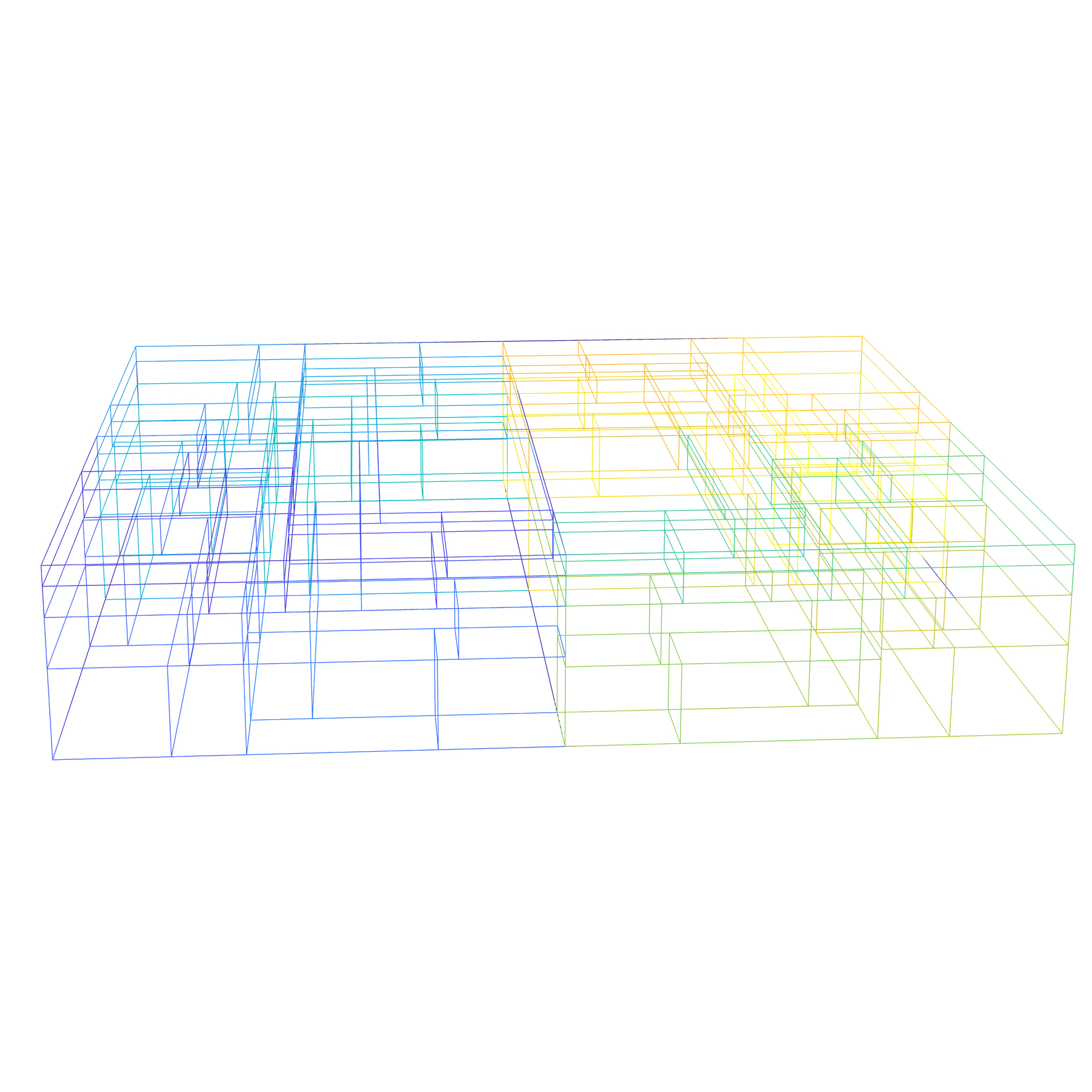}\label{SchematicKDTree}}
	\subfloat[b][R-tree]{\includegraphics[width=0.3\textwidth]{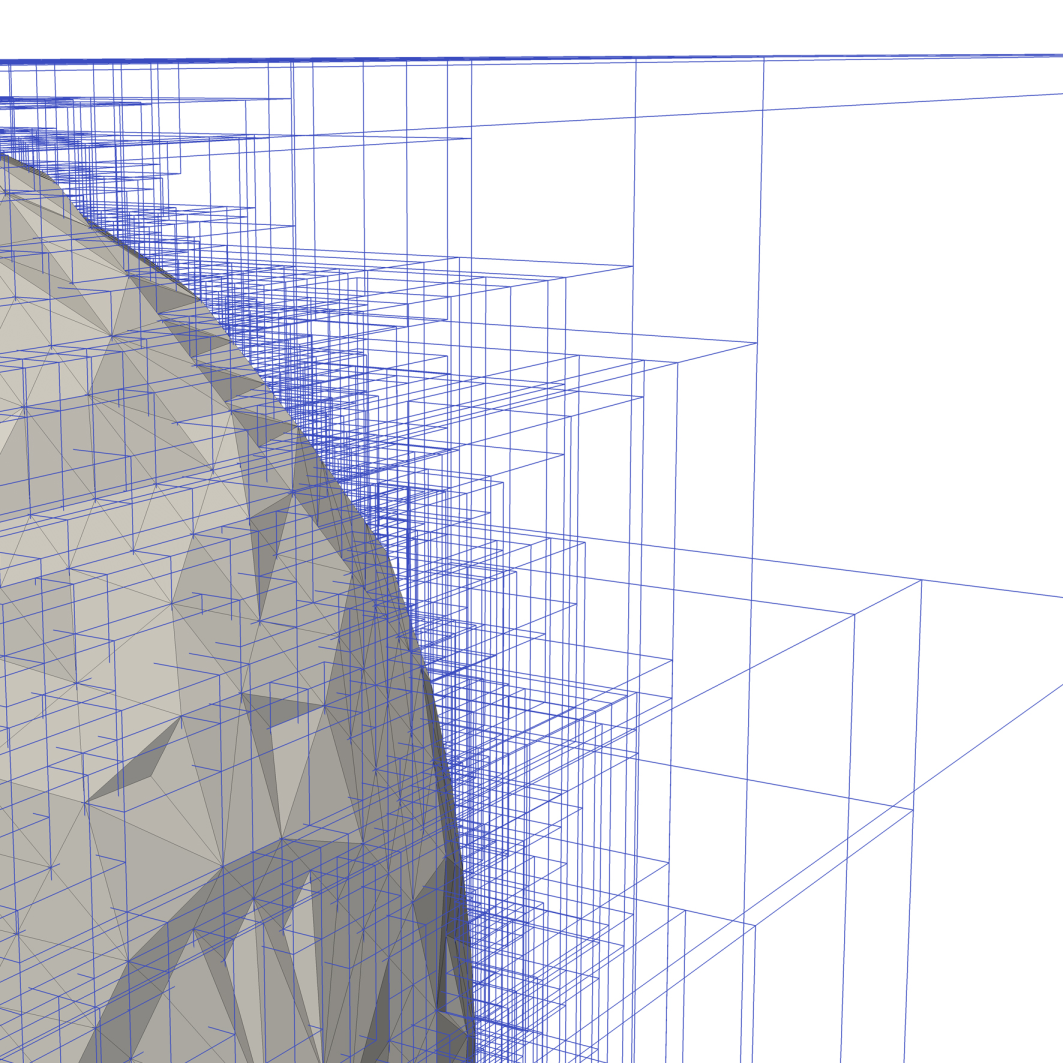}\label{SchematicRTree}}
	\subfloat[c][Spatial binning]{\includegraphics[width=0.30\textwidth]{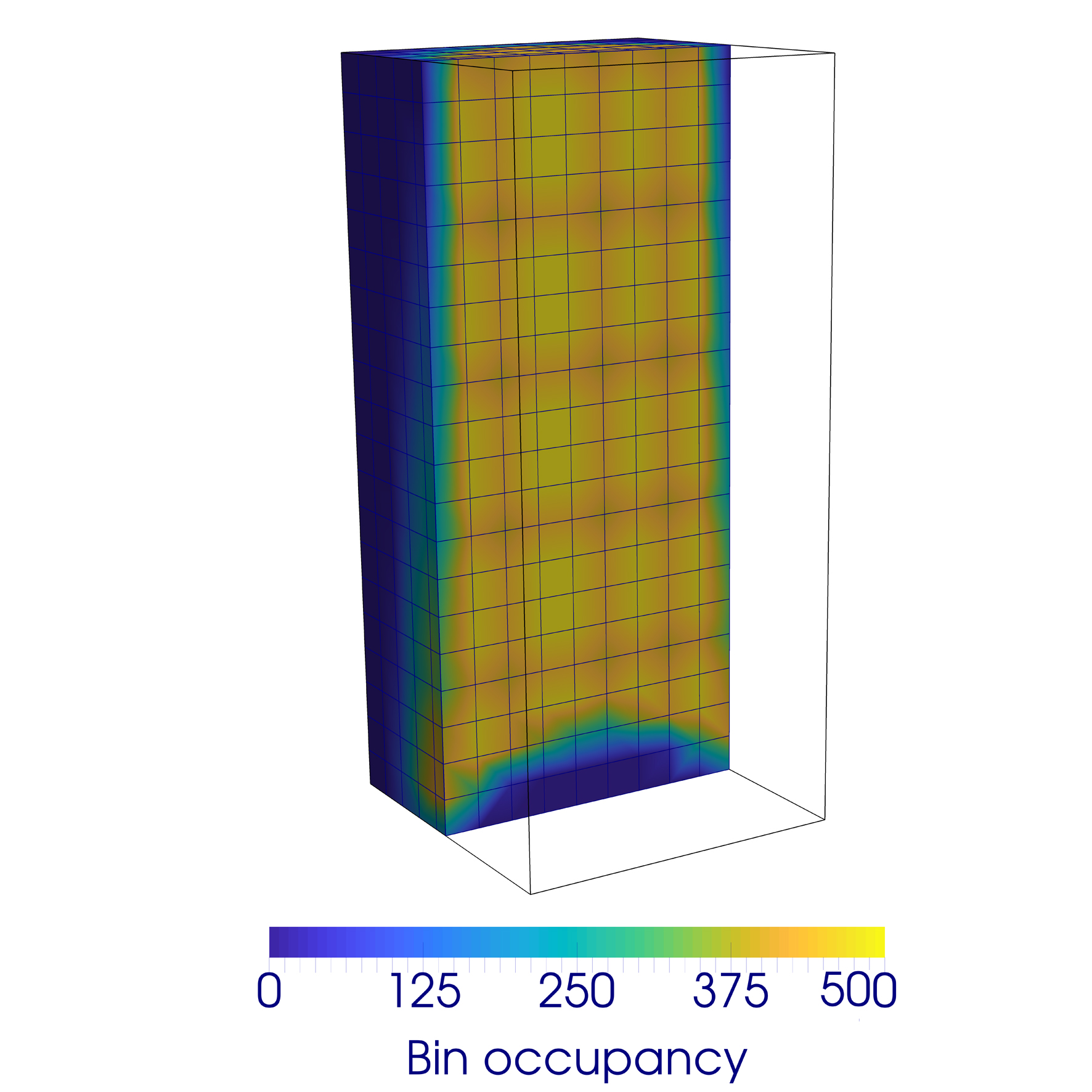}\label{SchematicBucketing}}
\caption{Different spatial indices were used to accelerate the filtering of the search space during querying: KD-trees a) realize a recursive bisectioning of the search space. The geometrical primitives (points) are grouped by their adjacency in the search space and stored in the so-called leafs of the tree. These leafs are organized in a hierarchy of boxes covering the search space. R-trees b) realize a similar recursive partitioning of the search space into a hierarchy of bounding boxes with additional data structures which allow the storage of triangles instead of points inside the leafs. Above tree structures enable queries with a logarithmic time complexity in the number of primitives. Alternatively, the search space c) can be divided regularly into a set of bins of fixed size and dimensions to offer near constant time queries. This approach is referred to as binning or gridding.}
\label{QueryingStructures}
\end{figure}

\paragraph{Details on efficient spatial indices}
Figure \ref{QueryingStructures} displays the types of spatial indices used for accelerating queries.

KD-trees (Fig. \ref{SchematicKDTree}) \cite{Bentley1975,Lang2010} have long been used within the APT community \cite{Haley2009,Felfer2013}. In our opinion, though, the topic is worth a deeper inspection for two reasons: first, the subtle modifications which are typically necessary to configure a spatial index for performance are seldom reported. Second, we observe that the utilization of parallelized spatial indices in the context of APT has neither been reported in substantial detail nor found implementation in most APT software. An example which supports the first reason pertains to the settings typically used for the leaf occupation, i.e. how many ions are packed into a leaf. Many users work with the \matlab{} KD-tree implementation and accept default values. However, these defaults should be carefully checked and always reported. Otherwise, the reporting of the analysis performance is not only difficult to compare but possibly also further performance enhancement could had been gained. A more efficient APT data mining algorithms can be implemented appyling scientific computing strategies \cite{Lang2010,Patwary2015,Patwary2016} such as a parallelized handling and optimized spatial indices.

Consequently, \paraprobe{} builds a collection of KD-trees per region, with one tree for every ion type. The leafs of the trees for the case studies in this paper were occupied with \SI[mode=text]{256}{} ions at most. Our KD-tree implementation re-packs the ions into linear arrays which store all ions of the same leaf in adjacent memory locations. Thereby, the queries show higher memory locality as compared to an implementation that would keep the storage order of the original evaporation sequence. For the benefit of a constant querying time, a regular binning (Fig. \ref{SchematicBucketing}) \cite{Goetz2015} was employed for the clustering tool. For ion-to-edge distancing \paraprobe{} works with an implementation of an R-tree (Fig. \ref{SchematicRTree}) that was originally developed for molecular dynamics simulations \cite{Hedges2017}. A global R-tree on the master thread was used.

\paragraph{Parallelized volume tessellations}
A substantial limitation of the hitherto explored strategies for tessellating large APT datasets has been to rely almost exclusively on \matlab{} and employing its \qhull{} \cite{Barber1996} library out of the box \cite{Haley2009,Felfer2015a,Felfer2015b,Breen2017}. As it is the case for the two other important libraries, namely \cgal{} \cite{CGAL2018} and \voroxx{} \cite{Rycroft2009}, however, this restricts practitioners to inherent sequential processing. Adding parallelism the conventional way, i.e. through modifying the source code of the library is a very tedious and error-prone strategy: This requires in-depth knowledge of the algorithms and expertise to remain thread-safe and geometrically consistent. Enduring the feeding of large datasets to the sequential libraries is equally not practical because all above libraries by design hold the entire tessellation in main memory. 

Evidently, the number of duplicated ions in the guard zones should ideally be as low as possible. However, there is no optimal strategy to identify a thickness of the guard zones without knowing \textit{a priori} the shape of the resulting Voronoi cells. Therefore, we set a guard zone thickness of five times the average ion distance. \voroxx{} \cite{Rycroft2009} library functionalities were utilized to characterize the neighbors of each cell, whereby possibly incorrectly computed cells were identified. \voroxx{} employs a spatial binning to accelerate point queries, similar to the one shown in Fig. \ref{SchematicBucketing}. We found empirically that an occupation of five ions per bin defined our performance sweet spot \cite{Rycroft2009}.

Once tessellated, a post-processing is necessary to correct for edge effects. Namely, the Voronoi cells of ions at the edge of the point cloud make contact with the walls of the global box causing incorrect shapes. We identified with preliminary studies that probing exclusively for contact with the walls of the global box is insufficient, though, to filter out these edge cells and correct for all edge effects. Instead, \paraprobe{} evaluates the ion-to-edge distance. Based on this information, we filter out all Voronoi cells within closer than a threshold distance $d_{ero}$ to the edge (see Fig. 6 in the main paper).


\paragraph{\xdmf{} and visualization details}
\xdmf{} stands for eXtensible Data Model and Format \cite{XDMF2018}. Such a file is useful to describe the lean metadata about the HDF5 content. Consequently, \hdmf{} is a frequently used combination with visualization tools like Paraview \cite{Schroeder2006,Ayachit2015} and VisIt \cite{Childs2012}. Interfacing these files with Blender \cite{Blender2020} is possible via \python{} scripting. In this paper we visualized with Paraview and Python.

It is common practice when processing APT data to instruct multiple reconstructions and conduct multiple analyses before publishing an APT study. One could store all these analyses in the same HDF5 file. This has the disadvantage, though, that any small change of the file alters its binary representation, which makes hashing of the file difficult. Therefore, \paraprobe{} groups results of individual analysis tasks and writes eventually multiple \hdf{} files to organize the results of a workflow.


\paragraph{Implementation details}
We implemented \paraprobe{} as a collection of \cxx{} tools. For the case studies discussed herein, the software was compiled with the Intel Studio 2018 (v18.0.1) compiler using optimization (-O3 -march=native). The programs were linked against the corresponding Intel MPI, and \imk{} libraries. Additional third-party software was compiled with and linked into the code: \cgal{} (v4.11.3) \cite{CGAL2018,Da2018}, \voroxx{} (v0.4.6) \cite{Rycroft2009}, the \rapidxml{} file parser (v1.13) \cite{Kalicinski2006}, and \hdmf{} (v1.10.2) \cite{HDF52018}. 

All multithreaded studies were executed on an Ubuntu 16.04.4 LTS workstation. It was equipped with two $18$-core Xeon Gold 6150 CPUs with \SI[mode=text]{283}{\giga\byte} shared main memory per CPU socket. The machine and its local storage, a conventional hard disk drive (model HGST HUH721010ALE600), were used exclusively. Analyses were executed as a single MPI process spawning $1$, $6$, or $36$ \omp{} threads, respectively. The mapping of the threads to the physical cores was orchestrated to facilitate localized memory utilization using the NUMA library \cite{Kleen2004}. Analyses employed single precision floating point arithmetics. \cgal{} and \voroxx{} used at least double precision floating point arithmetics. 

The hybrid-parallelized studies were processed on the TALOS computer cluster. We used at most $80$ of its computing nodes. Each node is equipped with two $20$-core Xeon Gold 6138 CPUs. Each node offers \SI[mode=text]{192}{\giga\byte} main memory. On TALOS analyses with one, ten, or $80$ nodes were instructed. The nodes were used exclusively with one MPI process per node which spawned $40$ OpenMP threads. Further details to the TALOS system are given elsewhere \cite{Kuehbach2020a}.

\begin{figure}[!ht]
\centering
	\includegraphics[width=1.0\textwidth]{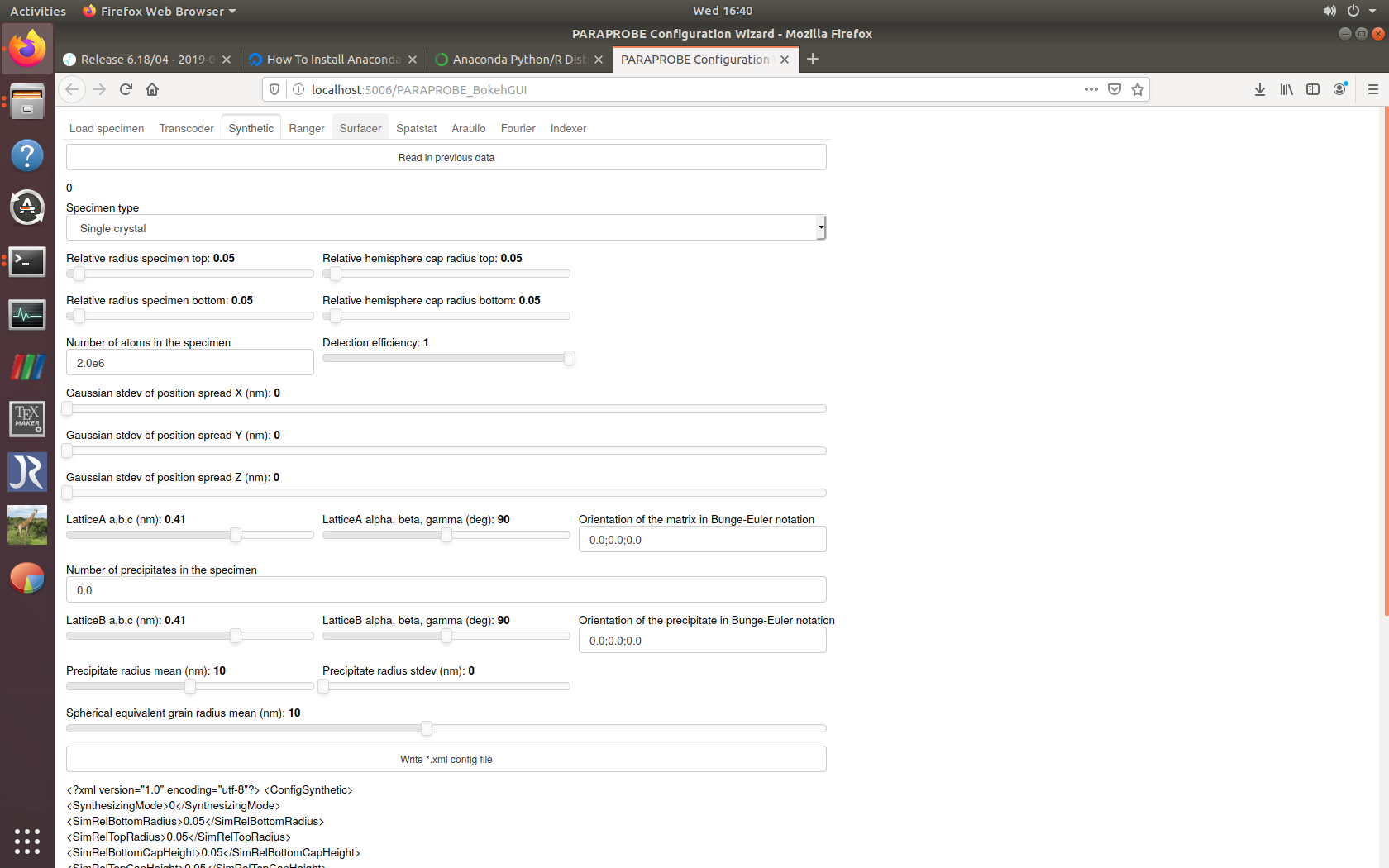}
\caption{A web browser-based \python{}/\bokeh{} app is available to assist users with configuring \paraprobe{} jobs. The GUI is under development. Here, it is exemplified the configuring of a synthetic specimen with the paraprobe-synthetic tool.}
\label{VisBokehGUI}
\end{figure}

\paragraph{Graphical user interface}
Figure \ref{VisBokehGUI} shows an example of a preliminary version of web browser-based graphical user interface with which \paraprobe{} settings files can be created for individual tools.

\section{References}
\sloppy
\printbibliography{}
\fussy

\end{document}